\documentclass[sigconf]{acmart}

\AtBeginDocument{%
  }

\copyrightyear{2026}
\acmYear{2026}
\setcopyright{cc}
\setcctype{by-nc-nd}
\acmConference[ICSE-SEIP '26]{2026 IEEE/ACM 48th International Conference on Software Engineering}{April 12--18, 2026}{Rio de Janeiro, Brazil}
\acmBooktitle{2026 IEEE/ACM 48th International Conference on Software Engineering (ICSE-SEIP '26), April 12--18, 2026, Rio de Janeiro, Brazil}
\acmPrice{}
\acmDOI{10.1145/3786583.3786898}
\acmISBN{979-8-4007-2426-8/2026/04}

\usepackage{hyperref}
\usepackage{url}
\usepackage{enumitem}
\usepackage{subcaption}

\usepackage{multirow}
\usepackage{pifont}
\usepackage{listings}
\usepackage{enumitem}
\usepackage{amsfonts}
\usepackage{makecell}

\begin{document}

\title{TerraFormer: Automated Infrastructure-as-Code with LLMs Fine-Tuned via Policy-Guided Verifier Feedback}

\author{Prithwish Jana}
\email{pjana7@gatech.edu}
\orcid{0000-0003-1967-4665}
\affiliation{%
  \institution{Georgia Institute of Technology}
  \city{Atlanta}
  \country{USA}
}
\authornote{Work carried out while affiliated to Amazon Web Services, Seattle, USA.}

\author{Sam Davidson}
\email{ssdavid@amazon.com}
\orcid{0000-0003-1160-2683}
\affiliation{%
  \institution{Amazon Web Services}
  \city{Seattle}
  \country{USA}
}

\author{Bhavana Bhasker}
\orcid{0009-0006-9605-5525}
\affiliation{%
  \institution{Amazon Web Services}
  \city{Seattle}
  \country{USA}
}

\author{Andrey Kan}
\orcid{0000-0002-2432-0047}
\affiliation{%
  \institution{Amazon Web Services}
  \city{Seattle}
  \country{USA}
}

\author{Anoop Deoras}
\orcid{0009-0007-4566-8767}
\affiliation{%
  \institution{Amazon Web Services}
  \city{Seattle}
  \country{USA}
}

\author{Laurent Callot}
\orcid{0000-0002-0756-8120}
\affiliation{%
  \institution{Amazon Web Services}
  \city{Seattle}
  \country{USA}
}



\renewcommand{\shortauthors}{Jana et al.}

\begin{abstract}
Automating Infrastructure-as-Code (IaC) is challenging, and large language models (LLMs) often produce incorrect configurations from natural language (NL). We present \textit{TerraFormer}, a neuro-symbolic framework for IaC generation and mutation that combines supervised fine-tuning with verifier-guided reinforcement learning, using formal verification tools to provide feedback on syntax, deployability, and policy compliance. We curate two large, high-quality NL-to-IaC datasets, \textsc{TF-Gen} (152k instances) and \textsc{TF-Mutn} (52k instances), via multi-stage verification and iterative LLM self-correction. Evaluations against 17 state-of-the-art LLMs, including $\sim$50$\times$ larger models like Sonnet 3.7, DeepSeek-R1, and GPT-4.1, show that \textit{TerraFormer} improves correctness over its base LLM by 15.94\% on IaC-Eval, 11.65\% on \textsc{TF-Gen} (Test), and 19.60\% on \textsc{TF-Mutn} (Test). It outperforms larger models on both \textsc{TF-Gen} (Test) and \textsc{TF-Mutn} (Test), ranks third on IaC-Eval, and achieves top best-practices and security compliance.

\end{abstract}

\ccsdesc[300]{Software and its engineering~Orchestration languages}
\ccsdesc[300]{Software and its engineering~Formal software verification}
\ccsdesc[300]{Computing methodologies~Natural language processing}
\ccsdesc[300]{Computing methodologies~Reinforcement learning}


\keywords{Infrastructure as Code (IaC), IaC generation, IaC mutation, Neuro-symbolic AI, Large language models, Formal Verification}


\maketitle

\section{Introduction}
\label{sec:intro}

Cloud computing has become the backbone of modern IT systems, with over 96\% of organizations using public cloud services and the market projected to exceed \$940 billion by 2026~\citep{spacelift2025stats}. Cloud-native development involves orchestrating diverse \textit{infrastructure} components such as compute, storage, networking, databases, and identity management, across multiple environments. To ensure reproducibility and support iterative changes, these \textit{resources} must be provisioned consistently and automatically. As a result, \textit{Infrastructure-as-Code (IaC)}~\citep{morris2020infrastructure} has emerged as a core DevOps practice, enabling infrastructure to be deployed and managed via high-level, machine-readable code (\textit{configuration}). Broadly, IaC tools fall into two categories: \textit{configuration management systems} (e.g., Puppet, Chef, Ansible) that manage already-provisioned systems' state, and \textit{infrastructure provisioning frameworks} (e.g., CloudFormation, Terraform, Pulumi) that declaratively define, provision, and manage resources. Among these, Terraform~\citep{terraform} is one of the most widely used IaC tools on GitHub, with its declarative \textit{HashiCorp Configuration Language (HCL)} experiencing over 30\% growth every year~\citep{octoverse2023,octoverse2024}. Backed by a strong community, Terraform’s modular, platform-agnostic definitions and plugin system enable integration with major cloud providers like AWS, Azure, and Google Cloud.


Although Terraform's HCL offers a readable, declarative syntax, authoring IaC configurations requires deep knowledge of cloud provider APIs, provider-specific attributes, and complex inter-resource dependencies. Experienced engineers typically spend around 100 minutes creating a moderately complex multi-cloud setup~\citep{ragothaman2024optimizing}, and human-written configurations often contain critical security vulnerabilities~\citep{buhler2025terrads}. These challenges highlight the need to automate IaC generation from high-level goals to reduce manual effort and security risks. Large language models (LLMs) are well-suited, having shown strong code generation across benchmarks. Notable examples include Claude 3.7~\citep{anthropic2025claude37}, GPT-4.1~\citep{openai2025gpt41}, DeepSeek-R1~\citep{guo2025deepseek}, LLaMA-4~\citep{meta_llama4_multimodal_2025}, and Qwen-3~\citep{yang2025qwen3}, which rank highly on HumanEval+~\citep{evalplus}, MMLU-Pro~\citep{wang2024mmlu}, Open LLM v2~\citep{open-llm-leaderboard-v2}, and Polyglot~\citep{aider_polyglot_2024}. However, applying LLMs to IaC automation involves unique challenges.

\begin{figure}[!t]
    \centering
    \begin{minipage}{0.95\linewidth}
        \includegraphics[width=\linewidth]{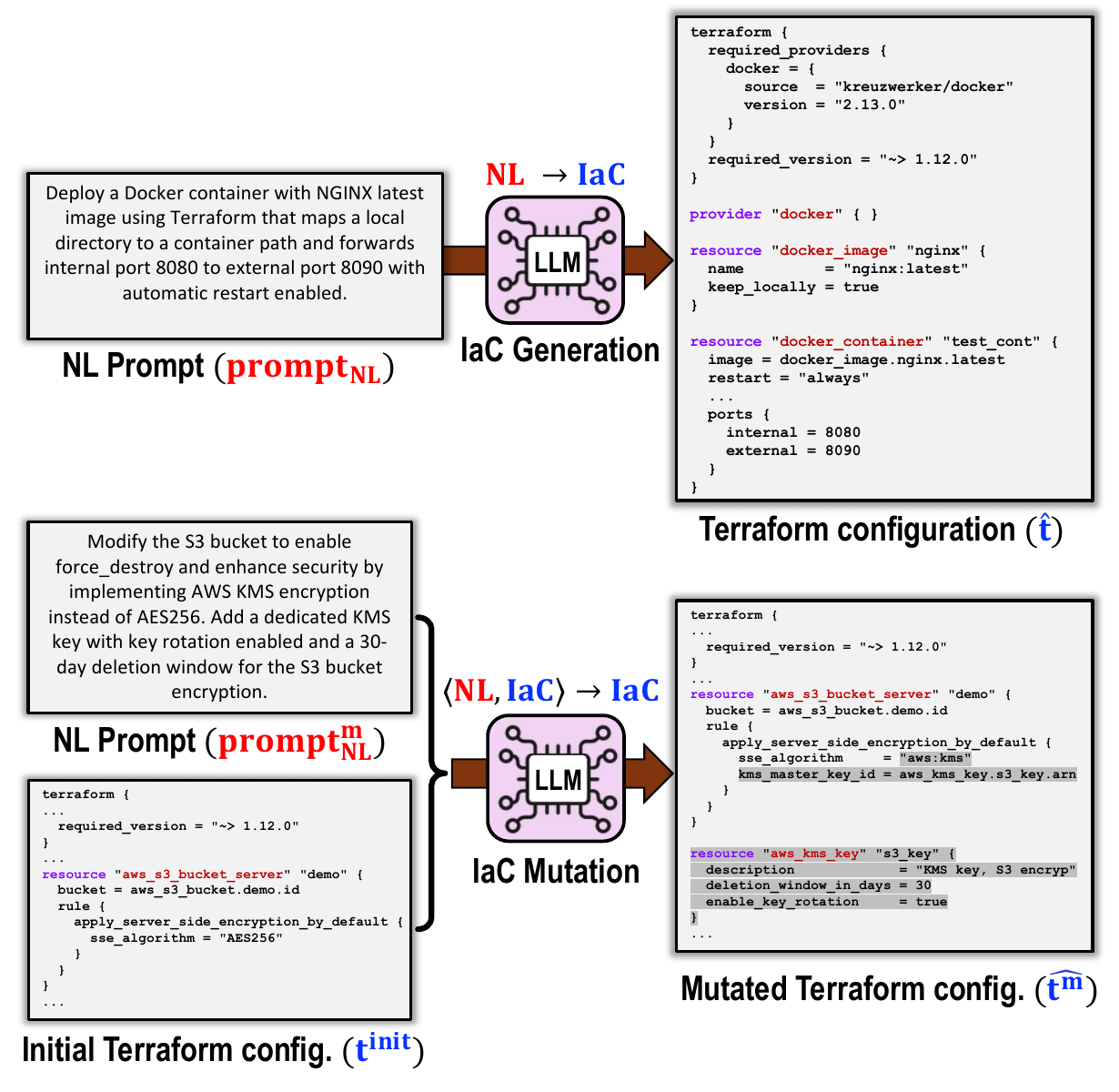}
    \end{minipage}
    \vspace{-2mm}
    \caption{\textsc{Learning Objectives.} Task definitions for IaC generation, i.e., creating Terraform configurations from scratch, and IaC mutation, i.e., modifying existing configurations, both taking natural language (NL) prompts as input.}
    \Description{Learning objectives showing task definitions for Infrastructure as Code generation, i.e., creating Terraform configurations from scratch, and Infrastructure as Code mutation, i.e., modifying existing configurations, both taking natural language prompts as input.}
    \label{fig:GenMutnPipeline}
\end{figure}

In this paper, we address two fundamental tasks in IaC automation using LLMs, focusing on Terraform and its native language HCL: (a) \textbf{IaC generation}, creating configurations from scratch given natural language (NL) prompts, and (2) \textbf{IaC mutation}, modifying existing configurations based on NL instructions (Figure~\ref{fig:GenMutnPipeline}). These use cases pose several challenges that limit the effectiveness of current LLMs in the IaC domain. 
\emph{\textbf{First}}, there is a lack of high-quality labeled datasets, hindering model evaluation and the development of specialized ones. Due to privacy restrictions, real-world IaCs are rarely shared, limiting LLMs’ exposure to diverse conventions and usage patterns in production.  
\emph{\textbf{Second}}, general-purpose LLMs struggle to satisfy the strict syntactic and semantic constraints of formal languages~\citep{jana2024neurosymbolic, ugare2024syncode} like HCL, frequently producing hallucinated or invalid code. This stems from their stochastic nature and is compounded by the lack of domain-specific supervision. The challenge is amplified in IaC, where tools like Terraform use a highly declarative, stateful language, and correctness depends on both syntax and semantic consistency with cloud provider APIs and inter-resource dependencies.
\emph{\textbf{Third}}, IaC mutation poses a distinct challenge from IaC generation, as it requires LLMs to accurately interpret existing infrastructure and make precise updates while preserving critical dependencies. Although a few recent studies~\citep{zhang2025deployability,palavalli2024using,kon2024iac,lee2024llm} have explored NL-to-IaC generation from scratch, the task of mutating or incrementally updating existing IaC via NL prompts (arguably more reflective of real-world development) remains largely underexplored.
\emph{\textbf{Fourth}}, automated evaluation of IaC correctness is a major bottleneck. Actual cloud deployment is prohibitively slow~\citep{qiu2023simplifying}, cannot reliably verify user intent, and often requires credentials unavailable to the LLM, necessitating manual post-processing. Pre-deployment testing remains an open problem~\citep{sokolowski2024automated}, so prior IaC research relies on proxies like BLEU, exact match~\citep{ganesh2023survey}, CodeBERTScore, LLM judges~\citep{kon2024iac}, linters~\citep{palavalli2024using}, and static analyzers~\citep{buhler2025terrads}. Without formal verifiers, these provide minimal guarantees for deployability or preservation of user intent.


To address the challenges of reliable IaC generation and mutation from NL prompts, we present \emph{TerraFormer}, a neuro-symbolic framework that leverages an LLM fine-tuned with feedback from formal verification tools. We employ a comprehensive suite of verifiers that assess Terraform IaC for compilability, deployability, and policy compliance. Using these verifiers, we curate a high-quality NL-to-IaC dataset by processing a large collection of erroneous Terraform configurations from open-source repositories through multi-stage verification. Crucially, those failing verification enter a \textit{multi-turn repair loop}, where an LLM iteratively refines them using verifier-generated error certificates. This self-correction process yields a large set of verified IaC, from which we derive NL prompts and policies that capture infrastructure intent. Though focused on Terraform, our automated data curation pipeline extends to other IaC tools (e.g., Pulumi, Ansible, CloudFormation) and general code generation or mutation. We use this dataset to fine-tune open-source LLMs for IaC generation and mutation. We warm-start with supervised fine-tuning (SFT), then apply reinforcement learning (RL) guided by fine-grained verifier feedback with progressively structured rewards, enabling the LLM to move beyond pattern imitation and produce deployable, functionally correct IaC.

\vspace{2mm}
\noindent\textbf{Contributions:}
\begin{itemize}[left=0em, topsep=0em]
    \item We present an automated data curation pipeline leveraging LLMs and formal verification to construct large-scale NL-to-IaC datasets: \textsc{TF-Gen} (152,475 generation instances; 330$\times$ IaC-Eval~\citep{kon2024iac}, 1000$\times$ DPIaC-Eval~\citep{zhang2025deployability}, 150$\times$ CloudEval-YAML~\citep{xu2024cloudeval}) and \textsc{TF-Mutn} (52,516 mutation instances, the first IaC mutation dataset), both with more complex infrastructures than IaC-Eval. We validate prompt–IaC–policy semantic alignment and mutation complexity via a cloud-expert survey, and curate test sets \textsc{TF-Gen} (Test) and \textsc{TF-Mutn} (Test) with 704 and 740 instances. (Sections~\ref{sec:dataset_construction},~\ref{sec:dataset_quality})
    \item We develop a training pipeline to create \textit{TerraFormer} for IaC generation and mutation, fine-tuning Qwen2.5-Coder (3B and 14B) on \textsc{TF-Gen} and \textsc{TF-Mutn}. Following an SFT warm-start, we apply RL with a fine-grained verifier-based reward that incorporates syntax, deployability, and policy compliance. (Section~\ref{sec:training})

    \item We extensively evaluate \textit{TerraFormer} against 17 SoTA LLMs, drawn from six model families (Anthropic, DeepSeek, Llama, OpenAI, Mistral, Qwen), on IaC generation and mutation, measuring correctness, deployability, syntactic validity, best practices, and security compliance. \textit{TerraFormer} significantly improves correctness over its base LLM (+15.94\% on IaC-Eval, +11.65\% on \textsc{TF-Gen} (Test), +19.60\% on \textsc{TF-Mutn} (Test)). It outperforms $\approx$50$\times$ larger foundation LLMs such as Sonnet 3.7, DeepSeek-R1, and GPT-4.1 on \textsc{TF-Gen} (Test) and \textsc{TF-Mutn} (Test), while achieving third-best performance on IaC-Eval. It also leads in best practices and security compliance. (Section~\ref{sec:expt}) 
\end{itemize}

\section{Related Work}
\label{sec:rel_work}

\textbf{LLM-based IaC Generation for Terraform.} Despite Terraform's prominence in DevOps, LLM-based NL-to-Terraform generation remains underexplored, and no prior work has addressed the task of mutation. \citet{ganesh2023survey} evaluate GPT-3.5-turbo (59\% accuracy) and CodeParrot-110M (8\%) on a small, unreleased 49-task AWS benchmark, using exact-match correctness against human-authored configurations. The challenges of Terraform generation are more evident on real-world datasets such as IaC-Eval~\citep{kon2024iac}, which consists of 458 human-curated, AWS-specific examples. To our knowledge, it is the only publicly available benchmark for this task. Contemporary LLMs perform poorly on IaC-Eval; the best-performing model, GPT-4, achieves under 20\% pass@1 accuracy. This aligns with~\citet{grant2024status}, who reports that SoTA LLMs struggle with Terraform synthesis, often hallucinating resource types and attribute names due to the relative scarcity of Terraform's HCL in LLM training data.

\noindent\textbf{LLM-based IaC Generation in Popular Formats (YAML/JSON).} In contrast to Terraform, LLMs perform better on IaC tools that use widely adopted general-purpose data formats like YAML and JSON, such as Ansible, Kubernetes, and AWS CloudFormation, owing to their fewer syntactic elements and greater presence in training data~\citep{grant2024status}. For instance, \citet{pujar2023automated} pre-trained CodeGen-350M-Multi on YAMLs from BigQuery and fine-tuned it on Ansible Galaxy data, raising functional correctness from 54.5\% (few-shot) to 70.8\%, though evaluation was limited by a 1024-token context. \citet{xu2024cloudeval} benchmarked 13 models on CloudEval-YAML for 1011 NL-to-Kubernetes tasks, with GPT-4-Turbo at 56.7\% functional correctness and LLaMA-2-70B-Chat leading among open-source models at 8.9\%. On DPIaC-Eval~\citep{zhang2025deployability}, Claude-3.5 achieved 30.2\% deployment success on 153 NL-to-CloudFormation tasks, rising to 95.5\% with 15 rounds of human feedback. Yet such feedback is often noisy and unreliable, yielding just 25.2\% intent satisfaction and poor security compliance.



\noindent\textbf{Neuro-Symbolic Code Generation.} Prior works on LLM-based formal language generation, such as code or proofs, leveraged symbolic feedback from compilers~\citep{liu2023rltf,dou2024stepcoder}, static analysis~\citep{jana2023attention}, and formal reasoning systems like LEAN~\citep{xin2024deepseek,jana2025proofbridge}, typically via RL. However, in data-scarce domains like Terraform IaC, existing methods rely on zero- or few-shot prompting without domain-specific supervision~\citep{delorenzo2025abstraction}. This gap stems from a lack of high-quality data and effective feedback; cloud deployments provide only Boolean signals, yielding sparse and inefficient rewards~\citep{jha2025rlsf}, and are slow. While linters improve NL-to-CloudFormation generation~\citep{palavalli2024using}, LLMs still struggle with errors and introduce new ones, exposing their hallucination tendencies and the limitations of linters as weak verifiers. 



\vspace{-1mm}
\section{Preliminaries and Problem Formulation}
\label{sec:prelim}



\begin{figure}[t]
    \fcolorbox{gray}{white}{%
      \includegraphics[width=0.26\textwidth]{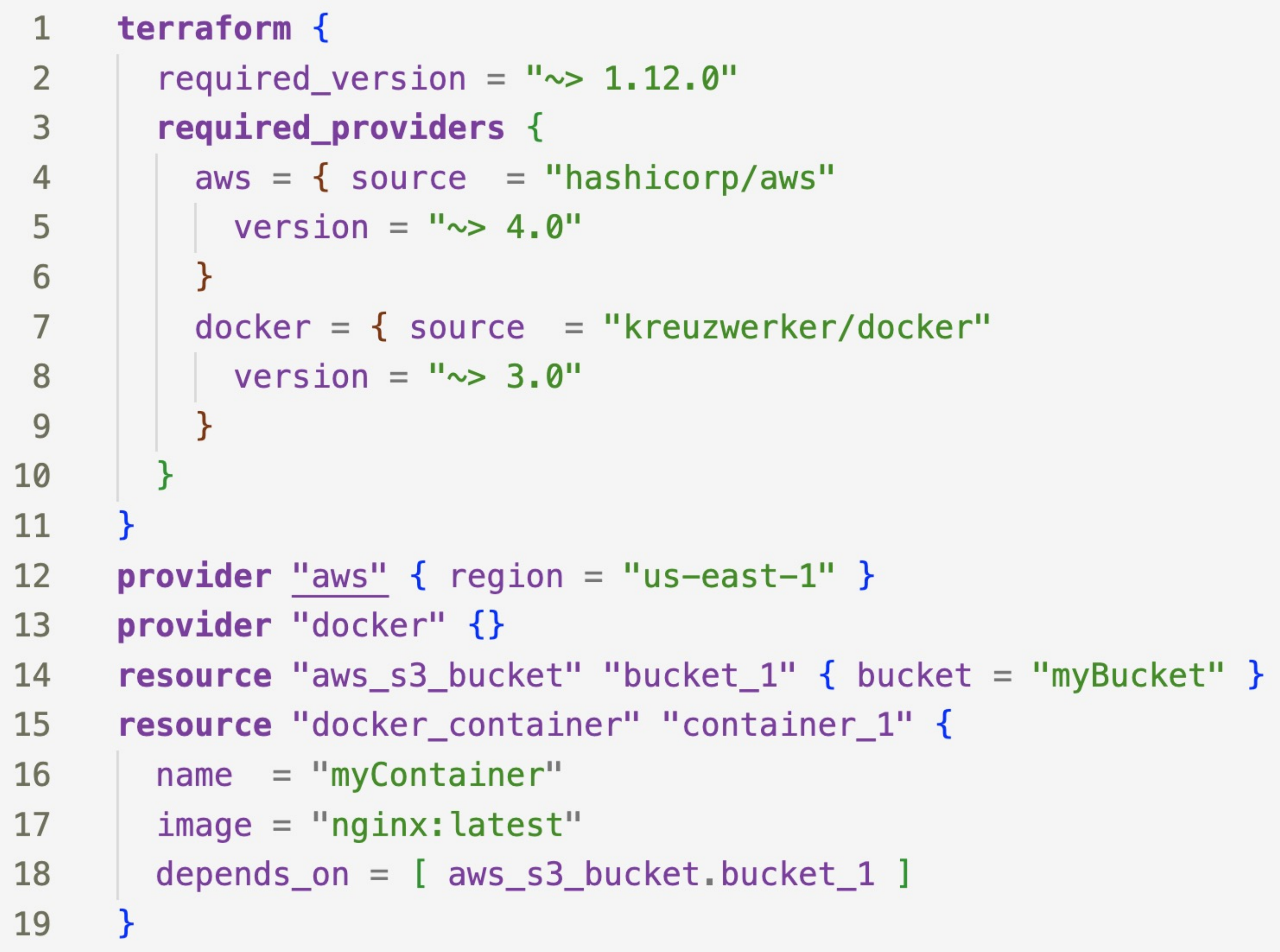}%
    }
  \vspace{-3mm}
  \caption{\textsc{Terraform Example.} IaC with 2 providers (AWS, Docker) and
  2 resources (S3, container) with a dependency.}
  \Description{Terraform example showing an Infrastructure as Code configuration with two providers, AWS and Docker, and two resources, an S3 bucket and a container, with a dependency between them.}
  \label{fig:terraformEg}
\end{figure}

Terraform configurations are written in HCL, a declarative domain-specific language (DSL) for specifying infrastructure state via \textit{blocks}. Key block types include \emph{provider} (e.g., \texttt{aws}, \texttt{azurerm}, \texttt{google}) for target clouds and \emph{resource} (e.g., \texttt{aws\_s3\_bucket}) for infrastructure components. Terraform supports complex multi-provider, multi-resource setups, allowing cross-cloud deployments. Inter-resource relations are expressed implicitly via references (e.g., a resource using another's attributes) or explicitly with \texttt{depends\_on}, to guide execution order. Figure~\ref{fig:terraformEg} shows a Terraform configuration example.


\noindent We define three \textit{formal verification oracles} for Terraform: \textbf{\scriptsize FV-i} checks \textit{compilability} (syntactic correctness), \textbf{\scriptsize FV-ii} verifies \textit{deployability} (semantic validity), and \textbf{\scriptsize FV-iii} ensures \textit{correctness} (policy compliance). \textbf{\scriptsize FV-i} and \textbf{\scriptsize FV-ii} are implemented via Terraform commands from~\citet{terraform} whereas, \textbf{\scriptsize FV-iii} is implemented by Open Policy Agent (OPA)~\citep{openpolicyagent}. Note that each stage is strictly stronger than the previous: correctness requires deployability, which requires compilability.

\begin{enumerate}[label={\scriptsize \textbf {[FV-\roman*]}}, left=-0.7em, itemsep=-0.2em, topsep=0em]
    \item \textbf{\texttt{terraform validate}}: Given a Terraform configuration, it performs static checks for \textit{syntax and structural consistency}. It ensures required fields and references are correctly defined, catching errors early before involving external systems.

    \item \textbf{\texttt{terraform plan}}: Given a configuration, it generates an execution DAG simulating deployment. It assesses \textit{semantic correctness} and \textit{runtime feasibility} by validating provider compatibility, resource dependencies, and runtime constraints.

    \item \textbf{\texttt{opa eval}}: We express infrastructure intent as machine-verifiable policies in Rego~\cite{openpolicyagent}, a declarative language for specifying rules over structured data like execution DAGs. A Rego policy acts as a unit-test for the DAG from \textbf{\scriptsize FV-ii}. Given this DAG and a policy, the oracle checks \textit{policy compliance}, verifying conformance with user intent, security best practices, organizational standards, and regulatory constraints. This adds a formal verification layer beyond syntax and deployability.
    
    
    
\end{enumerate}

\noindent Crucially, \textbf{\scriptsize FV-i}, \textbf{\scriptsize FV-ii}, and \textbf{\scriptsize FV-iii} produce detailed \textit{error certificates} upon failure, aiding diagnosis and debugging. Leveraging these certificates, we apply the oracles for two purposes. First, we construct large-scale verified datasets (\textsc{TF-Gen}, \textsc{TF-Mutn}) by passing configurations through the oracles, enabling a leading commercial-scale LLM to iteratively repair errors via inference-time feedback. Second, we use these oracles in RL-based fine-tuning of open-source LLMs, where the error certificates are converted into fine-grained reward signals to guide learning. In the following sections, we address two core learning objectives (Figure~\ref{fig:GenMutnPipeline}) in IaC automation using LLMs, with a focus on Terraform and its native declarative language, HCL:

\textbf{IaC Generation as a Learning Problem.} Given an NL prompt ($\mathit{{prompt}_{\text{NL}}}$), the goal is to generate a Terraform configuration ($\hat{t}$) by learning a function $f_{\text{gen}}(\mathit{{prompt}_{\text{NL}}}) \mapsto \hat{t}$ such that $\hat{t}$ is (a) compilable, (b) deployable, and (c) aligned with the intent of $\mathit{{prompt}_{\text{NL}}}$.

\textbf{IaC Mutation as a Learning Problem.} Given an existing Terraform configuration ($t^{\text{init}}$) and an NL prompt ($\mathit{{prompt}^{\text{m}}_{\text{NL}}}$), the goal is to generate an updated configuration ($\widehat{t^{\text{m}}}$) by learning $f_{\text{mutn}}(t^{\text{init}},\allowbreak \mathit{{prompt}^{\text{m}}_{\text{NL}}}) \mapsto \widehat{t^{\text{m}}}$ such that $\widehat{t^{\text{m}}}$ is (a) compilable, (b) deployable, and (c) faithful to the modification intent of $\mathit{{prompt}^{\text{m}}_{\text{NL}}}$.

    

\begin{figure*}[!t]
    \centering
    \begin{subfigure}[b]{0.95\textwidth}
        \includegraphics[width=\linewidth]{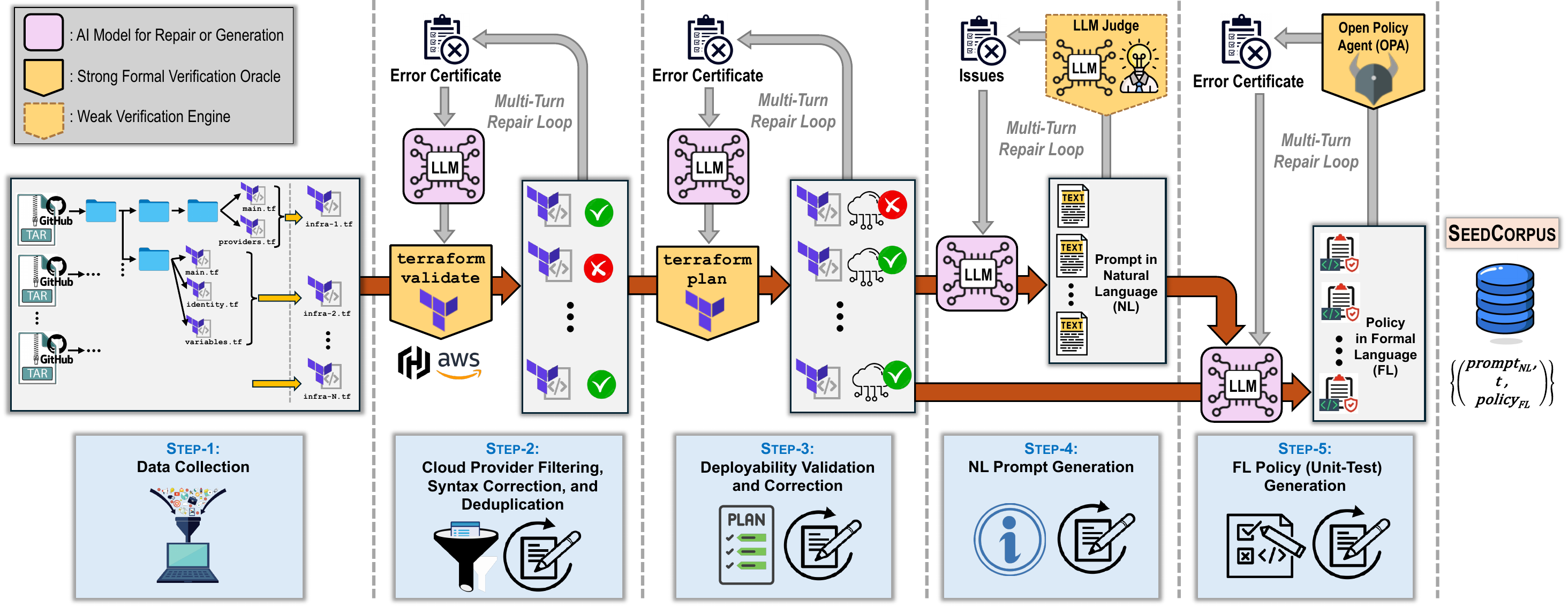}
        \vspace{-2.5mm}
        \caption{Pipeline to construct the \textsc{SeedCorpus}}
        \Description{Pipeline to construct the SeedCorpus}
        \label{fig:archDiagram-seed}
    \end{subfigure}
    \vspace{-2mm}
    \begin{subfigure}[b]{0.48\linewidth}
    \begin{center}
    \fcolorbox{gray!40}{white}{
      \makebox[\linewidth][c]{%
        \vbox{%
          \vspace{1pt}%
          \includegraphics[width=\linewidth]{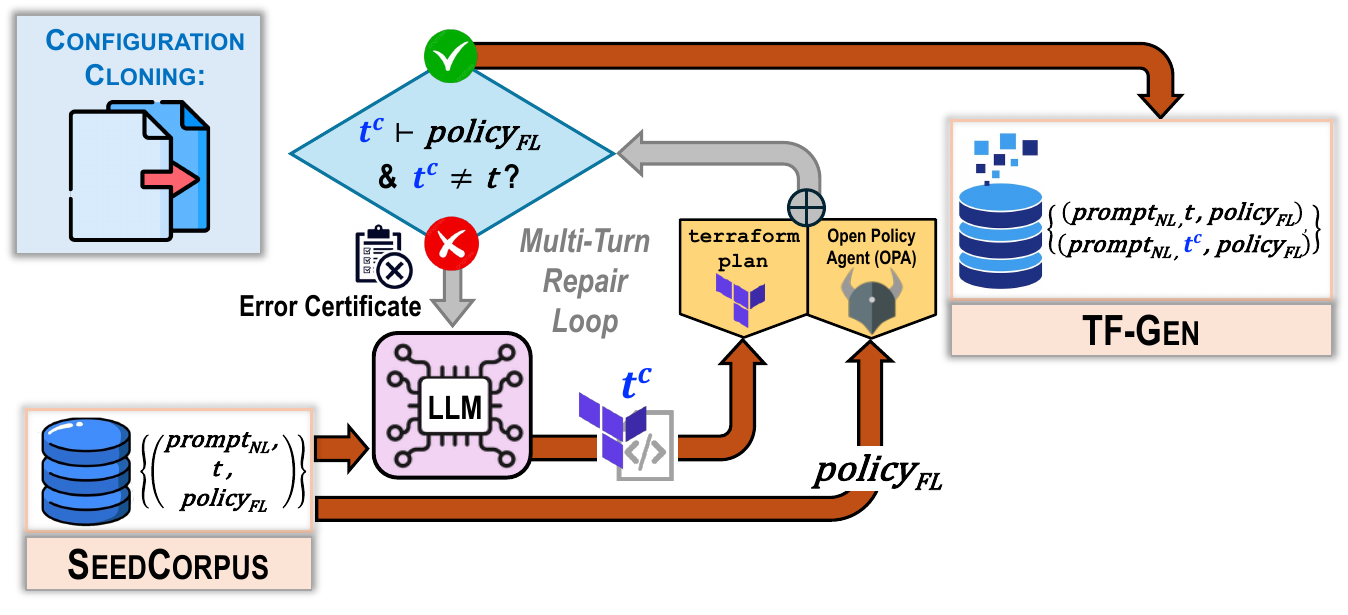}%
          \vspace{-2pt}%
        }%
      }
    }
    \end{center}
        \caption{Curation of \textsc{TF-Gen} dataset from \textsc{SeedCorpus}}
        \Description{Curation of TF-Gen dataset from SeedCorpus}
        \label{fig:archDiagram-gen}
    \end{subfigure}
    \hfill
    \begin{subfigure}[b]{0.5\linewidth}
    \begin{center}
    \fcolorbox{gray!40}{white}{
      \makebox[\linewidth][c]{%
        \vbox{%
          \vspace{1pt}%
          \includegraphics[width=\linewidth]{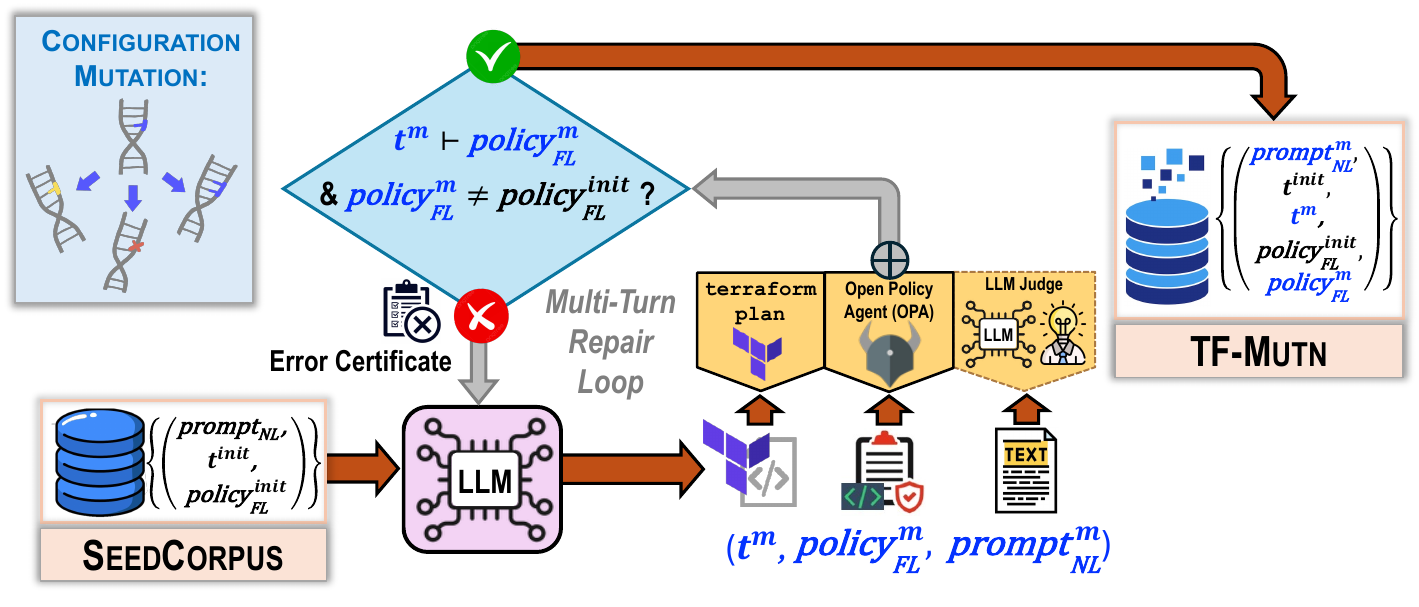}
          \vspace{-10pt}%
        }%
      }
    }
    \end{center}
        \caption{Curation of \textsc{TF-Mutn} dataset from \textsc{SeedCorpus}}
        \Description{Curation of TF-Mutn dataset from SeedCorpus}
        \label{fig:archDiagram-mutn}
    \end{subfigure}
\vspace{-1mm}
\caption{\textsc{Dataset Curation Pipeline.}
(a) Multi-turn repair loops with LLMs and verifiers produce verified Terraform configurations, NL prompts, and FL policies.
(b) \textsc{TF-Gen} extends \textsc{SeedCorpus} by cloning triplets to produce multiple implementations for an infrastructure.
(c) \textsc{TF-Mutn} augments \textsc{SeedCorpus} with mutated configurations, mutation prompts, and FL policies.}
\Description{Dataset curation pipeline. (a) Multi-turn repair loops using large language models and verifiers produce verified Terraform configurations, natural language prompts, and formal language policies. (b) TF-Gen extends SeedCorpus by cloning triplets to produce multiple implementations for an infrastructure. (c) TF-Mutn augments SeedCorpus with mutated configurations, mutation prompts, and formal language policies.}

    \label{fig:archDiagram}
    \vspace{1mm}
\end{figure*}

\section{\textsc{TF-Gen} and \textsc{TF-Mutn}: A Systematic Dataset Curation Pipeline with LLMs and Verification}
\label{sec:dataset_construction}

We present an automated pipeline for constructing datasets for formal language (e.g., code) generation and mutation tasks based on NL prompts. Using this framework, we create two comprehensive datasets: \textsc{TF-Gen} and \textsc{TF-Mutn}, for training and evaluating LLMs on \textit{IaC Generation} and \textit{IaC Mutation} tasks respectively. In both datasets, each infrastructure scenario includes a target Terraform configuration that is verified as deployable, along with an FL policy that the target configuration satisfies. The FL policy acts as a unit-test (or verifier function), enabling reliable, automated evaluation of LLM-generated IaC. We summarize both datasets below:

\begin{itemize}[left=0em, itemsep=0em, topsep=0em]
    \item \textsc{TF-Gen} includes \textit{152,475 infrastructure scenarios}, each structured as a triplet $\{\mathit{{prompt}_{\text{NL}}}, t, \mathit{{policy}_{\text{FL}}}\}$, comprising: (i) an NL prompt, (ii) a target Terraform configuration, and (iii) a FL policy codifying the constraints expected to hold in the target configuration.

    \item \textsc{TF-Mutn} includes \textit{52,516 infrastructure scenarios}, each represented as a quintuplet $\{\mathit{{prompt}^{\text{m}}_{\text{NL}}}, t^{\text{init}}, t^{\text{m}}, \mathit{{policy}^{\text{init}}_{\text{FL}}}, \mathit{{policy}^{\text{m}}_{\text{FL}}}\}$, consisting of: (i) an NL prompt, (ii) an initial Terraform configuration, (iii) a modified configuration (target) that implements the changes, and (iv-v) the initial and target FL policies.
\end{itemize}

\setlength{\tabcolsep}{3pt}
\begin{table*}[!t]
\centering
\caption{\textsc{Statistics for \textsc{TF-Gen} and \textsc{TF-Mutn}.} Datasets of IaC configurations in Terraform, showing min/median/max counts of providers, resources, configuration size (lines), and prompt length (words). IaC-Eval is included for comparison.}
\vspace{-2mm}
\label{tab:dataset-comparison}
\resizebox{0.82\textwidth}{!}{%
\begin{tabular}{lccr@{\hskip 20pt}
  *{2}{r}@{\hskip 10pt}  
  *{3}{r}@{\hskip 10pt}  
  *{3}{r}@{\hskip 10pt}  
  *{3}{r}@{\hskip 10pt}  
  *{3}{r}                
}
\toprule
\multirow{4}{*}{\textbf{Dataset}} & 
\multirow{4}{*}{\textbf{Task}} & 
\multirow{4}{*}{\textbf{Split}} & 
\multirow{4}{*}{\shortstack{\textbf{\# Source} \\ \textbf{Github} \\\textbf{Repos.}}} &
\multicolumn{2}{c}{\textbf{\# Instances}} &
\multicolumn{9}{c}{\textbf{Terraform Configuration Stats}} &
\multicolumn{3}{c}{\textbf{NL Prompt Stats}} \\
\cmidrule(lr){5-6} \cmidrule(lr){7-15} \cmidrule(lr){16-18}
& & & &
\textbf{\small\phantom{aa}Total} & \multirow{2}{*}{\small\shortstack{\textbf{\%with} \\ \textbf{policy}}} & 
\multicolumn{3}{c}{\textbf{\small \# Providers}} &
\multicolumn{3}{c}{\textbf{\small \# Resources}} &
\multicolumn{3}{c}{\textbf{\small \# Lines of Code}} &
\multicolumn{3}{c}{\textbf{\small Length (words)}} \\
\cmidrule(lr){7-9} \cmidrule(lr){10-12} \cmidrule(lr){13-15} \cmidrule(lr){16-18}
& & & & & & 
\textbf{\phantom{a}\small Min} & \textbf{\small Med} & \textbf{\small Max} & 
\textbf{\phantom{a}\small Min} & \textbf{\small Med} & \textbf{\small Max} & 
\textbf{\phantom{a}\small Min} & \textbf{\small Med} & \textbf{\small Max} & 
\textbf{\phantom{a}\small Min} & \textbf{\small Med} & \textbf{\small Max} \\
\midrule
\midrule
\multirow{1}{*}{\shortstack{
  IaC-Eval~\citep{kon2024iac}
}} & \multirow{1}{*}{IaC Generation} & \multirow{1}{*}{Full} & \multirow{1}{*}{N/A} & \multirow{1}{*}{458} & \multirow{1}{*}{100\%} & \multirow{1}{*}{0} & \multirow{1}{*}{1} & \multirow{1}{*}{4} & \multirow{1}{*}{0} & \multirow{1}{*}{3} & \multirow{1}{*}{27} & \multirow{1}{*}{3} & \multirow{1}{*}{30} & \multirow{1}{*}{294} & \multirow{1}{*}{3} & \multirow{1}{*}{17} & \multirow{1}{*}{215} \\
\midrule
\multirow{2}{*}{\textbf{TF-Gen (Ours)}} & \multirow{2}{*}{IaC Generation} & Train & 27,980 & 151,771  & 83\% & 0 & 1 & 35 & 0 & 4 & 199 & 1 & 111 & 4,496 & 10 & 44 & 134 \\
 & & Test & 704 & 704 & 100\% & 0 & 1 & 3 & 0 & 6 & 44 & 1 & 138 & 814 & 10 & 46 & 105 \\
\midrule
\multirow{2}{*}{\textbf{TF-Mutn (Ours)}} & \multirow{2}{*}{IaC Mutation} & Train & 19,427 & 51,776 & 100\% & 0 & 1 & 18 & 0 & 4 & 59 & 1 & 107 & 934 & 14 & 63 & 229 \\
 & & Test & 740 & 740 & 100\% & 0 & 1 & 5 & 0 & 5 & 41 & 1 & 134 & 704 & 18 & 66 & 190 \\
\bottomrule
\end{tabular}
}
\end{table*}

\noindent Table~\ref{tab:dataset-comparison} presents a statistical comparison with IaC-Eval~\citep{kon2024iac}, which to our knowledge is the only benchmark for NL-to-Terraform generation (458 instances). Our proposed \textsc{TF-Gen} dataset is $\sim$330$\times$ larger and rigorously validated using formal verification tools. It also far surpasses datasets for other IaC tools, being 1000$\times$ larger than DPIaC-Eval~\citep{zhang2025deployability} for AWS CloudFormation and 150$\times$ CloudEval-YAML~\citep{xu2024cloudeval} for Kubernetes. For IaC mutation task, \textsc{TF-Mutn} is the first dataset of its kind. Our datasets feature more complex infrastructure scenarios than IaC-Eval, with more providers, resources, and longer configurations on average (Table~\ref{tab:dataset-comparison}). An expert survey (Section~\ref{sec:dataset_quality}) on a subset of \textsc{TF-Gen} and \textsc{TF-Mutn} confirms overall quality. Together, we believe that the scale, complexity, and verification rigor of \textsc{TF-Gen} and \textsc{TF-Mutn} provide a strong foundation for advancing research in LLM-based IaC generation and mutation.



\textit{Modular, Extensible Dataset Curation Pipeline.} While our dataset curation pipeline is implemented for Terraform, it is readily adaptable to other IaC tools (e.g., Pulumi, Ansible) and general-purpose programming languages~\citep{jana2014cpp} (e.g., C++, Python) for curating similar code generation and mutation datasets. It involves simply swapping language-specific components, provided the target language has the notion of syntactic validity, semantic correctness, and testability. Moreover, the pipeline can accommodate future versions of Terraform by updating the associated formal verification oracles (listed in Section~\ref{sec:prelim}). Essentially, our pipeline relies on three key components: \emph{(a)} an initial corpus of code in the target language, possibly containing errors; \emph{(b)} a general-purpose LLM capable of reasoning about and generating code in that language; and \emph{(c)} syntax-aware compilers and formal verifiers that validate code both structurally and semantically, including against test cases. 

\textit{Stage-Wise Multi-Turn Repair Loops:} Our pipeline for curating \textsc{SeedCorpus} (Figure~\ref{fig:archDiagram-seed}) forms the basis of \textsc{TF-Gen} (Figure~\ref{fig:archDiagram-gen}) and \textsc{TF-Mutn} (Figure~\ref{fig:archDiagram-mutn}). Starting with an initial collection of Terraform configurations (step-1), many with errors, we verify them through three increasingly stringent stages: \textit{compilability} (step-2), \textit{deployability} (step-3), and \textit{policy compliance} (step-5). Crucially, verification is not used solely for filtering. At each stage, failed configurations enter an automated \textit{multi-turn repair loop}, where the verifier's diagnostic report (\textit{error certificate}) and the current configuration are given as in-context input to an LLM for revision (see prompt template in Appendix~\ref{sec-appendix:prompt_description}). The loop runs up to five turns, with the verifier rechecking each revision and issuing an updated certificate, so both evolve together as feedback guides repairs toward the stage's verification criteria. We use a leading commercial-scale foundation LLM that demonstrates strong zero-shot IaC performance and can reason over in-context feedback.

\textbf{Data Collection.} We start with TerraDS~\citep{buhler2025terrads}, a large corpus of real-world Terraform configurations in HCL, collected from 62{,}406 GitHub repositories under permissive licenses. Each repository contains one or more \textit{modules}, defined as subdirectories with at least one \texttt{.tf} file. As HCL is declarative, we concatenate all \texttt{.tf} files within each module into a single configuration, yielding 279{,}410 unified configurations. But, TerraDS lacks built-in validation, and analysis using Checkov~\citep{checkov} revealed widespread misconfigurations and security issues~\citep{buhler2025terrads}. These raise concerns about the reliability of using such unverified configurations for training or evaluating LLMs. To address this, we apply formal verification oracles to progressively \textit{validate} and \textit{repair} configurations, then augment them with NL prompts and FL policies, as detailed below.


\textbf{Cloud Provider Filtering, Syntax Correction, and Deduplication.} Our cloud-agnostic pipeline supports all major providers, including AWS, Azure, and Google Cloud. In this work, we focus on AWS due to its wide adoption~\citep{synergy2024cloudq2} and abundance of well-documented public configurations. We retain those using AWS and other HashiCorp-supported utility providers (\textit{random}, \textit{null}, \textit{local}, \textit{template}, \textit{tls}, \textit{time}, \textit{external}, \textit{http}, \textit{archive}, \textit{docker}, \textit{terraform}), yielding 117{,}520 configurations (42\% of the step-1 corpus). Syntax validation is then applied via our multi-turn repair loop using \textbf{\scriptsize FV-i} (Section~\ref{sec:prelim}) as verifier. To handle variables requiring user input, the LLM inserts plausible defaults during repair. This yields 101{,}751 syntax-validated configurations, 58\% repaired and the rest passing unchanged. After deduplication, we retain 97{,}645 unique configurations.




\textbf{Deployability Validation and Correction.} We next evaluate each syntax-validated configuration for deployability using \textbf{\scriptsize FV-ii} as the verifier, which checks for semantic correctness. Configurations failing this check undergo iterative fixes via the multi-turn repair loop guided by error certificates from \textbf{\scriptsize FV-ii}; those still failing after these attempts are discarded. This stage yields 89{,}843 deployable configurations, with $\sim$59\% recovered through automated repair.


\textbf{NL Prompt Generation.} Next, we invoke the LLM to generate NL prompts that allow for faithful reconstruction of each configuration. The LLM is instructed to begin each prompt with directive verbs (e.g., Generate, Set up, Deploy) and adopt a concise, goal-oriented tone that emphasizes \textit{what} to provision rather than \textit{how}. Prompt quality is refined\footnote{Determining whether an NL description is \textit{equivalent} to a formal language (FL) object is generally undecidable. However, generating NL from FL (\textit{informalization}) is usually tractable and sufficiently accurate~\citep{jiang2024multi} as FL objects contain rich semantic detail.} via a multi-turn repair loop in which an LLM judge compares the prompt to the Terraform code, provides structured feedback, and guides revisions until alignment is achieved. We provide prompts at three levels of abstraction: (i) high-level (broad overview), (ii) mid-level (moderate detail), and (iii) low-level (closely aligned with the configuration). Our mid-level prompts balance abstraction and detail to closely match human instructions, and thus we use them for all subsequent tasks.



\textbf{Formal Language Policy (Unit-Test) Generation.} For each deployable Terraform configuration and its NL prompt, we ask the LLM to generate an FL policy\footnote{LLMs generating test cases from code is an active research area~\citep{chen2024chatunitest, rahman2024automatic, zhang2025large}. To avoid trivial policies, we prompt the LLM to generate rule sets that cover all configuration elements in the NL prompt. An expert survey (Section~\ref{sec:dataset_quality}) confirms that the resulting multi-rule policies offer meaningful verification of each infrastructure scenario.} in Rego that captures the prompt's intent. The FL policy acts as a verification mechanism for the infrastructure. We validate each configuration against the generated policy using \textbf{\scriptsize FV-iii} (Section~\ref{sec:prelim}) and apply a multi-turn repair loop to iteratively refine policies that fail validation. From 89,843 configuration-prompt pairs, we generate valid policies for $\sim$71\%, yielding 63,576 triplets $\{\mathit{{prompt}_{\text{NL}}}, t, \mathit{{policy}_{\text{FL}}}\}$.



\textbf{\textsc{TF-Gen} Construction.} Since any $\mathit{{prompt}_{\text{NL}}}$ is inherently ambiguous, it can yield multiple configuration variants (\textit{clones}) that are semantically equivalent modulo $\mathit{{policy}_{\text{FL}}}$. To capture this diversity, we extend each triplet $\{\mathit{{prompt}_{\text{NL}}}, t, \mathit{{policy}_{\text{FL}}}\}$ in \textsc{SeedCorpus}, by prompting the LLM to generate a configuration clone $t^c$. We apply our multi-turn repair loop to $t^c$, using \textbf{\scriptsize FV-ii} and \textbf{\scriptsize FV-iii} (Section~\ref{sec:prelim}) to verify that it is deployable, satisfies $\mathit{{policy}_{\text{FL}}}$, and is distinct from $t$ (Figure~\ref{fig:archDiagram-gen}). We generate at most one clone per configuration, resulting in 153{,}175 IaC generation tasks in \textsc{TF-Gen}.


\begin{figure}[!t]
    \centering
    \begin{subfigure}[b]{\linewidth}
        \begin{subfigure}[b]{\linewidth}
            \begin{center}
            \fcolorbox{gray!40}{white}{
              \makebox[\linewidth][c]{%
                \vbox{%
                  \vspace{2pt}%
                  \includegraphics[width=0.7\linewidth]{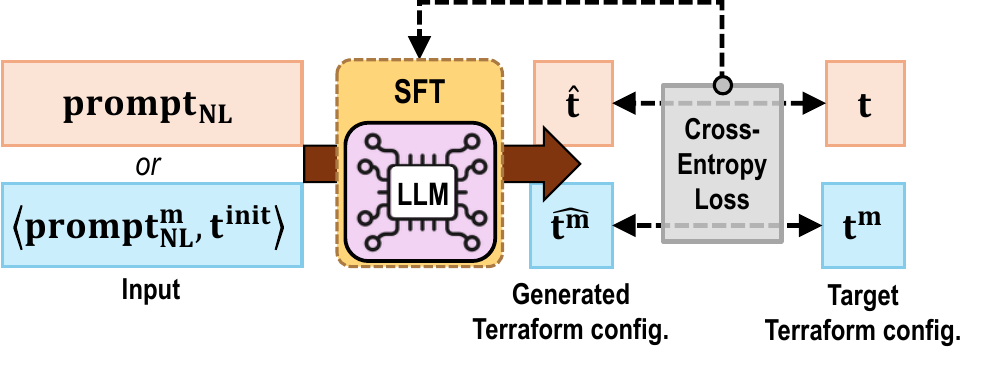}%
                }%
              }
            }
            \end{center}
            \caption{SFT via cross-entropy loss of generated and target configurations}
            \label{fig:SFT}
        \end{subfigure}
        \begin{subfigure}[b]{\linewidth}
            \begin{center}
            \fcolorbox{gray!40}{white}{
              \includegraphics[width=\linewidth]{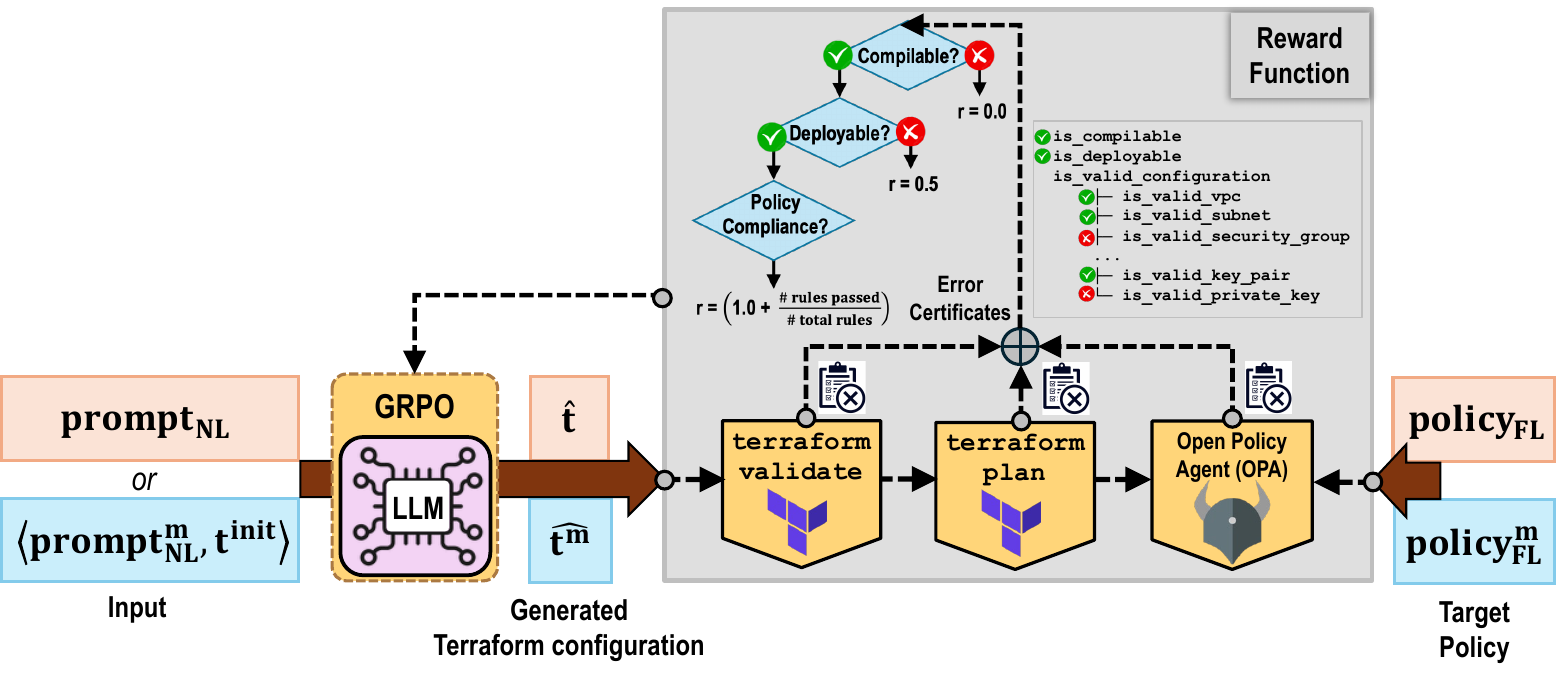}
            }
            \end{center}
            \caption{RL via fine-grained verifier feedback from \textbf{\scriptsize FV-i}, \textbf{\scriptsize FV-ii}, and \textbf{\scriptsize FV-iii}}
            \label{fig:RL}
        \end{subfigure}
    \end{subfigure}
    \vspace{-7mm}
    \caption{\textsc{Learning Objectives.} Instruction-tuning LLMs for the tasks of IaC generation and mutation via SFT and RL}
    \Description{Learning objectives for instruction-tuning large language models for Infrastructure as Code generation and mutation. The top subfigure shows supervised fine-tuning using cross-entropy loss between generated and target Terraform configurations. The bottom subfigure shows reinforcement learning using fine-grained verifier feedback from multiple formal verifiers.}

    \vspace{2mm}
    \label{fig:sftrl}
\end{figure}

\textbf{\textsc{TF-Mutn} Construction.} For $\{\mathit{{prompt}_{\text{NL}}}, t^{\text{init}}, \mathit{{policy}^{\text{init}}_{\text{FL}}}\}$ in \textsc{SeedCorpus}, we prompt the LLM to produce a mutated configuration $t^{\text{m}}$, an updated policy ($\mathit{policy}^{\text{m}}_{\text{FL}}$) and an NL prompt ($\mathit{prompt}^{\text{m}}_{\text{NL}}$) describing the mutation. Mutations alter or add resources, providers, or variables, while keeping all artifacts consistent. A multi-turn repair loop with \textbf{\scriptsize FV-ii} and \textbf{\scriptsize FV-iii} ensures $t^{\text{m}}$ is deployable and satisfies $\mathit{policy}^{\text{m}}_{\text{FL}}$, while confirming $\mathit{policy}^{\text{m}}_{\text{FL}}$ differs from $\mathit{{policy}^{\text{init}}_{\text{FL}}}$ (Figure~\ref{fig:archDiagram-mutn}). The loop also checks prompt faithfulness via an LLM judge. We generate at most one mutation per \textsc{SeedCorpus} instance, yielding 52,516 quintuplets $\{\mathit{{prompt}^{\text{m}}_{\text{NL}}},\allowbreak t^{\text{init}},\allowbreak t^{\text{m}},\allowbreak \mathit{{policy}^{\text{init}}_{\text{FL}}},\allowbreak \mathit{{policy}^{\text{m}}_{\text{FL}}}\}$. 

\textbf{Train/Test Splits.} To ensure out-of-distribution evaluation, we create test splits for both \textsc{TF-Gen} and \textsc{TF-Mutn} by randomly selecting instances that are traced back (during the \textit{Data Collection} step) to GitHub repositories containing exactly one module. This prevents related modules from the same repository appearing in both training and test sets, avoiding data leakage and enabling evaluation on unseen module contexts. Table~\ref{tab:dataset-comparison} summarizes the dataset statistics for these splits.

\section{Fine-Tuning Methodology}
\label{sec:training}

\textbf{Supervised Fine-Tuning (SFT).} For IaC generation, the LLM takes $\mathit{prompt}_{\text{NL}}$ as input and outputs configuration $\hat{t}$. For SFT, it is trained to minimize the cross-entropy loss between $\hat{t}$ and target $t$ (Figure~\ref{fig:SFT}):
\vspace{-4mm}

\begin{equation}
\mathcal{L}^{\theta}_{\text{CE}}(t, \hat{t}) = -\frac{1}{T} \sum_{j=1}^T \sum_{k=1}^{|V|} \mathbf{1}_{t_j = V_k} \times\log P_\theta(\hat{t}_j = V_k)
\end{equation}
\vspace{-2mm}

\noindent \noindent where $\theta$ are the trainable parameters, $T$ is the maximum tokenized length of $t$ and $\hat{t}$, $V$ is the vocabulary, $\mathbf{1}_{t_j = V_k}$ equals $1$ if the $j^{\text{th}}$ token of $t$ is the $k^{\text{th}}$ token in $V$ and $0$ otherwise, and $P_\theta(\hat{t}_j = V_k)$ is the probability that the model predicts the $k^{\text{th}}$ token of $V$ at position $j$ in $\hat{t}$. For the IaC mutation task, the LLM takes $\langle \mathit{prompt}^{\text{m}}_{\text{NL}}, t^{\text{init}} \rangle$ as input and outputs a mutated configuration $\widehat{t^m}$, minimizing $\mathcal{L}^{\theta}_{CE}(t^m, \widehat{t^m})$.

We ensure that SFT is effective for IaC generation and mutation as follows. First, in Section~\ref{sec:dataset_construction}, all target Terraform configurations in \textsc{TF-Gen} and \textsc{TF-Mutn} are verified as syntactically correct and deployable, and semantic alignment between each prompt and its target is checked using an LLM judge. This alignment is confirmed via a cloud-expert survey on a sample set (Section~\ref{sec:dataset_quality}), ensuring a high-quality training dataset. Second, through cloning, \textsc{TF-Gen} includes multiple diverse configurations for the same NL prompt, capturing the many valid ways to realize an infrastructure. All variants are formally verified to be equivalent modulo a policy that encodes the prompt’s infrastructure intent. This design enables the LLM to learn robust mappings from high-level prompts to valid, deployable configurations across a broad solution space.

\textbf{Reinforcement Learning (RL) via Policy-Guided Verifier Feedback.} Even with a high-quality dataset and increasing diversity through configuration cloning, SFT remains limited by its reliance on pattern imitation. To address this, we apply RL after an SFT warm-start. Training examples in \textsc{TF-Gen} (Train) and \textsc{TF-Mutn} (Train) have an associated target policy, making it suitable for RL-based fine-tuning. We ensure that this policy serves as a reliable verifier function by checking that the target configuration satisfies the policy and that semantic alignment between the prompt and its policy is validated via an LLM judge (Section~\ref{sec:dataset_construction}) and a cloud-expert survey (Section~\ref{sec:dataset_quality}); this verifier is then used to compute a fine-grained reward. In the RL setting for IaC generation, the LLM acts as a policy $\pi_\theta$, taking $\mathit{prompt}_{\text{NL}}$ as input and producing a configuration $\hat{t}$ as an action. 
This action is evaluated by verifiers \textbf{\scriptsize FV-i}, \textbf{\scriptsize FV-ii}, and \textbf{\scriptsize FV-iii} (Section~\ref{sec:prelim}) to compute a fine-grained reward $\in [0,2]$:

\vspace{-4mm}
\begin{equation}
r(\hat{t}, \mathit{{policy}_{\text{FL}}}) =
\begin{cases}
1 + \frac{\text{\# rules of}\;\mathit{{policy}_{\text{FL}}}\;\text{passed}}{\text{\# total rules in}\;\mathit{{policy}_{\text{FL}}}}, & \text{deployable},\\[1mm]
0.5, & \text{compilable},\\[1mm]
0, & \text{otherwise.}
\end{cases}
\label{eq:reward}
\end{equation}

\noindent Illustrated in Figure~\ref{fig:RL}, we fine-tune the LLM via Group Relative Policy Optimization (GRPO)~\cite{shao2024deepseekmath} using the fine-grained verifier reward, with a KL-divergence penalty~\citep{palenicek2021survey}, as follows:

\vspace{-4mm}
\begin{equation}
\begin{aligned}
\mathcal{L}^{\theta}_{\text{GRPO}}
&= \mathbb{E}_{\hat{t} \sim \pi_\theta(\cdot \mid \mathit{prompt}_{\text{NL}})} 
\Big[
\scriptstyle
\frac{\pi_\theta(\hat{t}\mid \mathit{prompt}_{\text{NL}})}
     {\pi_{\theta_{\text{SFT}}}(\hat{t}\mid \mathit{prompt}_{\text{NL}})}
\;\big( r(\hat{t}, \mathit{policy}_{\text{FL}}) - b \big)
\Big] \\
&\quad - \beta\times \, \text{KL}\!\left(\pi_\theta \;\|\; \pi_{\theta_{\text{SFT}}}\right),
\end{aligned}
\end{equation}

\noindent where $\pi_\theta$ is the current policy, 
$\pi_{\theta_{\text{SFT}}}$ the SFT reference, 
$r(\hat{t}, \mathit{policy}_{\text{FL}})$ the fine-grained verifier reward (Eq.~\ref{eq:reward}), 
$b$ the group-relative baseline, 
$\beta$ the KL penalty coefficient, and 
$\text{KL}(\pi_\theta \,\|\, \pi_{\theta_{\text{SFT}}})$ 
regularizes the policy toward the SFT model. The verifier feedback in Eq.~\ref{eq:reward} produces a gradual progression of reward, encouraging the model to iteratively generate better configurations through training, from syntactic correctness to deployability, and ultimately to functional correctness. For the IaC mutation task, the LLM takes $\langle \mathit{prompt}^{\text{m}}_{\text{NL}}, t^{\text{init}} \rangle$ as input and outputs a mutated configuration $\widehat{t^m}$, on which $r(\widehat{t^m}, \mathit{{policy}^{\text{m}}_{\text{FL}}})$ is computed. Crucially, the model is rewarded for syntactic correctness, deployability, and the extent of functional correctness, rather than for textual similarity to a finite set of target configurations.




\section{Experimental Results}
\label{sec:expt}

\setlength{\tabcolsep}{0pt}
\begin{table*}[!t]
\centering

\caption{\textsc{Evaluation of IaC Generation.} Comparison of SoTA LLMs and TerraFormer on IaC-Eval and \textsc{TF-Gen} (Test). TerraFormer* and TerraFormer\textsuperscript{\textdagger} are fine-tuned from Qwen2.5-Coder-3B (*) and -14B (\textdagger), respectively. Reported improvements (in parentheses) are relative to these base models. Bold and underline denote the best and second-best performance per column.}
\vspace{-2mm}
\label{tab:iac-gen}
\resizebox{\textwidth}{!}{%
\begin{tabular}{
  l@{\hskip 8pt}  
  l@{\hskip 0pt}  
  c@{\hskip 3pt} 
  @{\hskip 2pt}c@{\hskip 2pt}  
  *{5}{c@{\hskip 10pt}}  
  @{\hskip 2pt}c@{\hskip 2pt}  
  *{5}{c@{\hskip 10pt}}  
}
\toprule
\multicolumn{3}{c@{\hskip 0pt}}{\textbf{Tool/LLM}}
&  
\multicolumn{6}{c@{\hskip 10pt}}{\textbf{Evaluation on IaC-Eval}} &
&  
\multicolumn{5}{c}{\textbf{Evaluation on \textsc{TF-Gen} (Test)}} \\
\cmidrule(lr){1-3} \cmidrule(lr){5-9} \cmidrule(lr){11-15}
\multirow{2}{*}{\textbf{Method}} & 
\multirow{2}{*}{\textbf{Model}} & 
\multirow{2}{*}{\textbf{\#Params.}} &
&  
\makecell{\textbf{Correctness}\\\textbf{(\%)}} & 
\makecell{\textbf{Deployability}\\\textbf{(\%)}} & 
\makecell{\textbf{Compilability}\\\textbf{(\%)}} & 
\makecell{\textbf{Linter}\\\textbf{Pass Rate (\%)}} & 
\makecell{\textbf{Security}\\\textbf{Compliance (\%)}} &
& 
\makecell{\textbf{Correctness}\\\textbf{(\%)}} & 
\makecell{\textbf{Deployability}\\\textbf{(\%)}} & 
\makecell{\textbf{Compilability}\\\textbf{(\%)}} & 
\makecell{\textbf{Linter}\\\textbf{Pass Rate (\%)}} & 
\makecell{\textbf{Security}\\\textbf{Compliance (\%)}} \\
\midrule
\midrule
\multirow{17}{*}{\makecell[l]{\textbf{Few-shot}\\\textbf{inference}\\\textbf{of SoTA}\\\textbf{LLMs}}} & Claude Sonnet 3.7 & UNK & \vrule width 0.5pt & \textbf{35.37} & 70.09 & \textbf{76.64} & 99.34 & \underline{67.22} & \vrule width 0.5pt & \underline{16.90} & \textbf{78.69} & \textbf{85.23} & \underline{99.86} & 55.49 \\
& DeepSeek(DS)-R1 & 671B & \vrule width 0.5pt & \underline{33.84} & \underline{72.27} & \underline{75.55} & 99.13 & 63.67 & \vrule width 0.5pt & 11.65 & 61.93 & 79.69 & 98.72 & 53.41 \\
& OpenAI GPT 4.1 & UNK & \vrule width 0.5pt & 25.98 & 56.99 & 59.17 & \textbf{100.00} & 62.87 & \vrule width 0.5pt & 10.37 & 51.99 & 61.36 & \underline{99.86} & 52.79 \\
\cmidrule(lr){2-15}
& Llama-3.3-70B-Instruct & 70B & \vrule width 0.5pt & 20.96 & 50.00 & 51.75 & \textbf{100.00} & 66.59 & \vrule width 0.5pt & 7.95 & 40.63 & 49.43 & 98.15 & 53.50 \\
& DS-R1-Distill-Llama-70B & 70B & \vrule width 0.5pt & 13.54 & 32.31 & 35.59 & 99.34 & 61.22 & \vrule width 0.5pt & 7.10 & 23.86 & 31.53 & 99.43 & 53.84 \\
\cmidrule(lr){2-15}
& DS-Coder-33B-Instruct & 33B & \vrule width 0.5pt & 14.63 & 40.39 & 42.14 & 70.52 & 40.22 & \vrule width 0.5pt & 7.10 & 34.38 & 44.46 & 76.99 & 39.00 \\
\cmidrule(lr){2-15}
& Qwen2.5-Coder-14B\textbf{\textsuperscript{\textdagger}} & 14B & \vrule width 0.5pt & 15.50 & 42.58 & 43.89 & 96.94 & 58.89 & \vrule width 0.5pt & 6.39 & 30.54 & 40.20 & 93.32 & 48.24 \\
& Qwen3-14B & 14B & \vrule width 0.5pt & 11.14 & 25.98 & 30.35 & \underline{99.78} & 57.23 & \vrule width 0.5pt & 5.97 & 29.12 & 38.78 & \underline{99.86} & 52.64 \\
& CodeLlama-13B-Instruct & 13B & \vrule width 0.5pt & 0.22 & 7.64  & 36.03 & 96.29 & 36.14 & \vrule width 0.5pt & 3.55 & 20.17 & 33.95 & 93.61 & 44.44 \\
& Llama-3.2-11B-Instruct & 11B & \vrule width 0.5pt & 8.08 & 31.44 & 35.37 & 98.03 & 59.53 & \vrule width 0.5pt & 5.54 & 23.44 & 29.97 & 96.16 & 50.64 \\
\cmidrule(lr){2-15}
& Qwen3-8B & 8B & \vrule width 0.5pt & 4.80 & 14.63 & 18.12 & 98.91 & 52.42 & \vrule width 0.5pt & 4.97 & 17.47 & 23.01 & 98.01 & 49.98 \\
& DS-R1-0528-Qwen3-8B & 8B & \vrule width 0.5pt & 1.97 & 5.46 & 11.57 & 75.76 & 40.98 & \vrule width 0.5pt & 2.27 & 4.97 & 9.52 & 81.82 & 48.59 \\
& Llama-3.1-8B & 8B & \vrule width 0.5pt & 1.97 & 12.23 & 31.00 & 99.13 & 46.27 & \vrule width 0.5pt & 5.26 & 19.74 & 29.55 & 95.74 & 49.52 \\
& DS-R1-Distill-Llama-8B & 8B & \vrule width 0.5pt & 0.22 & 0.87 & 1.53 & 80.79 & 47.27 & \vrule width 0.5pt & 0.57 & 0.99 & 1.14 & 94.60 & 60.64 \\
& Mistral-7B-Instruct & 7B & \vrule width 0.5pt & 3.06 & 8.30 & 10.26 & 97.38 & 60.88 & \vrule width 0.5pt & 2.70 & 5.11 & 6.68 & 95.03 & \underline{65.68} \\
\cmidrule(lr){2-15}
& Qwen2.5-Coder-3B\textbf{*} & 3B & \vrule width 0.5pt & 6.55 & 18.78 & 20.96 & 97.38 & 57.39 & \vrule width 0.5pt & 4.55 & 18.18 & 20.45 & 94.89 & 48.14 \\
& Llama-3.2-3B-Instruct & 3B & \vrule width 0.5pt & 1.53 & 5.68 & 10.26 & 89.30 & 62.44 & \vrule width 0.5pt & 2.41 & 6.82 & 9.80 & \textbf{100.00} & 56.33 \\
\midrule
\multirow{4}{*}{\makecell[l]{\textbf{Our}\\\textbf{approach}}} & \textbf{TerraFormer* (SFT)} & \textbf{3B} & \vrule width 0.5pt & 12.45 {\footnotesize(+5.90)} & 36.96 {\footnotesize(+21.18)} & 43.23 {\footnotesize(+22.27)} & \textbf{100.00} {\footnotesize(+2.62)} & 62.12 {\footnotesize(+4.73)} & \vrule width 0.5pt & 8.95 {\footnotesize(+4.40)} & 40.62 {\footnotesize(+22.44)} & 44.89 {\footnotesize(+24.44)} & \textbf{100.00} {\footnotesize(+5.11)} & 61.79 {\footnotesize(+13.65)} \\
& \textbf{TerraFormer* (SFT+RL)} & \textbf{3B} & \vrule width 0.5pt & 19.43 {\footnotesize(+12.88)} & 44.10 {\footnotesize(+25.32)} & 48.03 {\footnotesize(+27.07)} & \textbf{100.00} {\footnotesize(+2.62)} & 64.29 {\footnotesize(+6.90)} & \vrule width 0.5pt & 10.51 {\footnotesize(+5.96)} & 49.01 {\footnotesize(+30.83)} & 53.13 {\footnotesize(+32.68)} & \textbf{100.00} {\footnotesize(+5.11)} & 62.98 {\footnotesize(+14.84)} \\
\cmidrule(lr){2-15}
 & \textbf{TerraFormer\textsuperscript{\textdagger} (SFT)} & \textbf{14B} & \vrule width 0.5pt & 24.23 {\footnotesize(+8.73)} & 56.99 {\footnotesize(+14.41)} & 60.70 {\footnotesize(+16.81)} & \textbf{100.00} {\footnotesize(+3.06)} & 65.00 {\footnotesize(+6.11)} & \vrule width 0.5pt & 11.93 {\footnotesize(+5.54)} & 56.96 {\footnotesize(+26.42)} & 61.08 {\footnotesize(+20.88)} & \textbf{100.00} {\footnotesize(+6.68)} & 58.11 {\footnotesize(+9.87)} \\
& \textbf{TerraFormer\textsuperscript{\textdagger} (SFT+RL)} & \textbf{14B} & \vrule width 0.5pt & 31.44 {\footnotesize(+15.94)} & \textbf{72.93} {\footnotesize(+30.35)} & 75.11 {\footnotesize(+31.22)} & \textbf{100.00} {\footnotesize(+3.06)} & \textbf{69.34} {\footnotesize(+10.45)} & \vrule width 0.5pt & \textbf{18.04} {\footnotesize(+11.65)} & \underline{70.17} {\footnotesize(+39.63)} & \underline{79.97} {\footnotesize(+39.77)} & \textbf{100.00} {\footnotesize(+6.68)} & \textbf{66.98} {\footnotesize(+18.74)} \\

\bottomrule
\end{tabular}%
}
\end{table*}

\begin{figure*}[!t]
    \centering
\begin{minipage}[t]{0.35\textwidth}
    \centering
    \begin{minipage}{\textwidth}
        \vspace{-10pt} 
        \includegraphics[width=\textwidth]{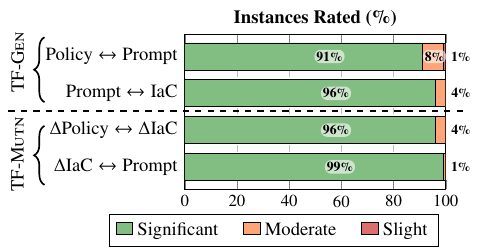}
        \includegraphics[width=\textwidth]{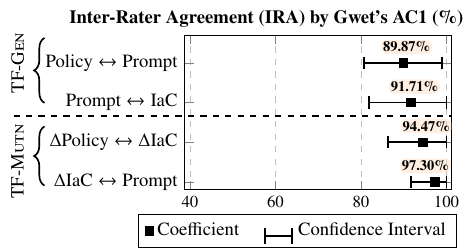}%
    \end{minipage}
    \vspace{-3mm}
    \caption{\textsc{Survey on Semantic Alignment.} Cloud-expert ratings for prompt–IaC–policy alignment on sampled \textsc{TF-Gen} and \textsc{TF-Mutn} instances, and inter-rater agreement.}
    \Description{Survey on semantic alignment showing cloud expert ratings for alignment between prompts, Infrastructure as Code configurations, and policies on sampled TF-Gen and TF-Mutn instances, along with inter-rater agreement.}

    \label{fig:humanAnnot}
\end{minipage}
    \hfill
    \begin{minipage}[t]{0.62\textwidth}
    \vspace{-29mm}
    \captionof{table}{\textsc{Evaluation of IaC Mutation.} Comparison of SoTA LLMs and TerraFormer on \textsc{TF-Mutn} (Test). Best per column bolded, 2nd-best underlined.}
    \vspace{-2mm}
        \centering
    \resizebox{\textwidth}{!}{%
    \begin{tabular}{
      l@{\hskip 8pt}  
      l@{\hskip 0pt}  
      c@{\hskip 3pt} 
      @{\hskip 2pt}c@{\hskip 2pt}  
      *{5}{c@{\hskip 10pt}}  
    }
    \toprule
    \multicolumn{3}{c@{\hskip 0pt}}{\textbf{Tool/LLM}}
    &  
    \multicolumn{6}{c}{\textbf{Evaluation on \textsc{TF-Mutn} (Test)}} \\
    \cmidrule(lr){1-3} \cmidrule(lr){5-9}
    \multirow{2}{*}{\textbf{Method}} & 
    \multirow{2}{*}{\textbf{Model}} & 
    \multirow{2}{*}{\textbf{\#Params.}} &
    &  
    \makecell{\textbf{Correctness}\\\textbf{(\%)}} & 
    \makecell{\textbf{Deployability}\\\textbf{(\%)}} & 
    \makecell{\textbf{Compilability}\\\textbf{(\%)}} & 
    \makecell{\textbf{Linter}\\\textbf{Pass Rate (\%)}} & 
    \makecell{\textbf{Security}\\\textbf{Compliance (\%)}} \\
    \midrule
    \midrule
    \multirow{17}{*}{\makecell[l]{\textbf{Few-shot}\\\textbf{inference}\\\textbf{of SoTA}\\\textbf{LLMs}}} 
    & Claude Sonnet 3.7 & UNK & \vrule width 0.5pt & \underline{52.03} & \textbf{93.51} & \textbf{94.73} & 99.86 & 59.78 \\
    & DeepSeek(DS)-R1 & 671B & \vrule width 0.5pt & 42.03 & 80.41 & 84.86 & 95.00 & 55.93 \\
    & OpenAI GPT 4.1 & UNK & \vrule width 0.5pt & 41.62 & 81.62 & \underline{84.46} & 99.32 & 60.21 \\
    \cmidrule(lr){2-9}
    & Llama-3.3-70B-Instruct & 70B & \vrule width 0.5pt & 38.38 & 77.70 & 81.08 & 99.86 & 60.64 \\
    & DS-R1-Distill-Llama-70B & 70B & \vrule width 0.5pt & 31.49 & 62.97 & 65.95 & 99.59 & 61.81 \\
    \cmidrule(lr){2-9}
    & DS-Coder-33B-Instruct & 33B & \vrule width 0.5pt & 28.78 & 63.92 & 67.57 & 94.46 & 56.75 \\
    \cmidrule(lr){2-9}
    & Qwen2.5-Coder-14B\textbf{\textsuperscript{\textdagger}} & 14B & \vrule width 0.5pt & 35.54 & 69.86 & 73.38 & 99.32 & 59.72 \\
    & Qwen3-14B & 14B & \vrule width 0.5pt & 30.95 & 55.14 & 57.30 & 99.59 & 59.47 \\
    & CodeLlama-13B-Instruct & 13B & \vrule width 0.5pt & 24.32 & 52.43  & 55.41 & 96.22 & 57.37 \\
    & Llama-3.2-11B-Instruct & 11B & \vrule width 0.5pt & 18.65 & 37.84 & 50.54 & 99.46 & 60.62 \\
    \cmidrule(lr){2-9}
    & Qwen3-8B & 8B & \vrule width 0.5pt & 26.49 & 47.43 & 49.19 & 99.05 & 59.57 \\
    & DS-R1-0528-Qwen3-8B & 8B & \vrule width 0.5pt & 20.81 & 38.24 & 40.00 & 89.46 & 53.65 \\
    & Llama-3.1-8B & 8B & \vrule width 0.5pt & 20.95 & 40.68 & 42.84 & 97.84 & 61.41 \\
    & DS-R1-Distill-Llama-8B & 8B & \vrule width 0.5pt & 13.92 & 21.35 & 22.03 & 98.78 & 63.86 \\
    & Mistral-7B-Instruct & 7B & \vrule width 0.5pt & 7.57 & 14.59 & 16.08 & 95.95 & 61.74 \\
    \cmidrule(lr){2-9}
    & Qwen2.5-Coder-3B\textbf{*} & 3B & \vrule width 0.5pt & 19.73 & 33.51 & 35.27 & 94.19 & 59.44 \\
    & Llama-3.2-3B-Instruct & 3B & \vrule width 0.5pt & 1.35 & 2.03 & 20.68 & 74.19 & 47.83 \\
    \midrule
    \multirow{4}{*}{\makecell[l]{\textbf{Our}\\\textbf{approach}}} 
    & \textbf{TerraFormer* (SFT)} & \textbf{3B} & \vrule width 0.5pt & 32.03 {\footnotesize(+12.30)} & 61.76 {\footnotesize(+28.25)} & 63.38 {\footnotesize(+28.11)} & 99.59 {\footnotesize(+5.40)} & 66.38 {\footnotesize(+6.94)} \\
    & \textbf{TerraFormer* (SFT+RL)} & \textbf{3B} & \vrule width 0.5pt & 37.97 {\footnotesize(+18.24)} & 68.11 {\footnotesize(+34.60)} & 70.00 {\footnotesize(+34.73)} & \underline{99.89} {\footnotesize(+5.70)} & \underline{67.21} {\footnotesize(+7.77)} \\
    \cmidrule(lr){2-9}
    & \textbf{TerraFormer\textsuperscript{\textdagger} (SFT)} & \textbf{14B} & \vrule width 0.5pt & 43.91 {\footnotesize(+8.37)} & 72.16 {\footnotesize(+2.30)} & 78.38 {\footnotesize(+5.00)} & \textbf{100.00} {\footnotesize(+0.68)} & 66.23 {\footnotesize(+6.51)} \\
    & \textbf{TerraFormer\textsuperscript{\textdagger} (SFT+RL)} & \textbf{14B} & \vrule width 0.5pt & \textbf{55.14} {\footnotesize(+19.60)} & \underline{82.57} {\footnotesize(+12.71)} & 83.92 {\footnotesize(+10.54)} & \textbf{100.00} {\footnotesize(+0.68)} & \textbf{67.89 }{\footnotesize(+8.17)}\\
    \bottomrule
    \end{tabular}%
    }
        \label{tab:iac-mutn}
    \end{minipage}
\end{figure*}

We evaluate a diverse suite of LLMs across two tasks: \textit{IaC generation} and \textit{IaC mutation}. For IaC generation (Table~\ref{tab:iac-gen}), we use the IaC-Eval dataset~\citep{kon2024iac}, which contains 458 instances, as well as our \textsc{TF-Gen} (Test) dataset with 704 instances. For IaC mutation (Table~\ref{tab:iac-mutn}), evaluation is performed on our \textsc{TF-Mutn} (Test) dataset, which comprises 740 instances. Dataset statistics are summarized in Table~\ref{tab:dataset-comparison}, with reproducibility details in Appendix~\ref{sec-appendix:reprod}.

\textbf{State-of-the-art (SoTA) Baselines.} As noted in Section~\ref{sec:rel_work}, there are currently no LLMs or tools specifically tailored for Terraform IaC. To fill this gap, our evaluation covers 17 LLMs across four categories: (i) \textbf{large-scale foundation models} with over 100B parameters: Claude Sonnet 3.7~\citep{anthropic2025claude37}, DeepSeek-R1~\citep{guo2025deepseek}, and GPT-4.1~\citep{openai2025gpt41}. We exclude Gemini~\citep{team2023gemini} due to subscription restrictions and its low Terraform performance (11.96\%) on IaC-Eval reported by~\citet{kon2024iac}, below GPT-3.5-turbo. (ii) \textbf{distilled variants of foundation models on smaller backbone LLMs}, including DeepSeek-R1-Distill-Llama-70B, DeepSeek-R1-0528-Qwen3-8B, DeepSeek-R1-Distill-Llama-8B~\citep{guo2025deepseek}, evaluated alongside their backbones Llama-3.3-70B-Instruct~\citep{meta_llama33_modelcard}, Qwen3-8B~\citep{yang2025qwen3} and Llama-3.1-8B~\citep{dubey2024llama}; (iii) \textbf{instruction-tuned LLMs} across varying sizes, such as Llama-3.3-70B-Instruct~\citep{meta_llama33_modelcard}, Llama-3.2-Instruct (11B and 3B)~\citep{metallama32blog}, and Mistral-7B-Instruct~\citep{jiang2023mistral7b}; and (iv) \textbf{code-specialized instruction-tuned LLMs}, including DeepSeek-Coder-33B-Instruct~\citep{guo2024deepseek}, Qwen2.5-Coder (14B and 3B)~\citep{hui2024qwen2}, and CodeLlama-13B-Instruct~\citep{roziere2023code}. All models are evaluated on few-shot inference, each given three in-context examples for IaC generation or mutation (prompts in Appendix~\ref{sec-appendix:prompt_description}). Overall, models evaluated span 3B-671B parameters. Sonnet 3.7, DeepSeek-R1, and Mistral models were run via Amazon Bedrock, GPT-4.1 via the OpenAI API, and other models from Huggingface were hosted on a server with 8$\times$40GB NVIDIA A100 GPUs.

\begin{figure*}[!t]
    \vspace{-2mm}  
    \centering

    \begin{subfigure}[t]{0.5\textwidth}
        \centering
        \includegraphics[height=2.8cm]{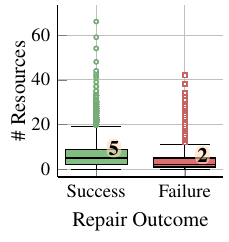}\hspace{2mm}%
        \includegraphics[height=2.8cm]{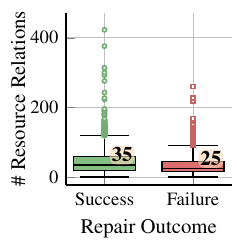}\hspace{2mm}%
        \includegraphics[height=2.8cm]{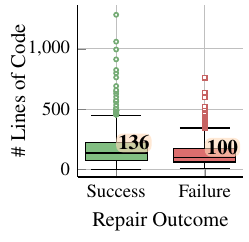}%

        \caption{Policy generation outcome vs. input IaC complexity}
        \label{fig:subfig1}
    \end{subfigure}%
    \begin{subfigure}[t]{0.5\textwidth}
        \centering
        \includegraphics[height=2.8cm]{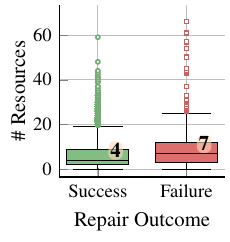}\hspace{2mm}%
        \includegraphics[height=2.8cm]{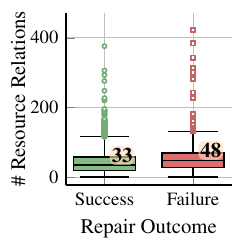}\hspace{2mm}%
        \includegraphics[height=2.8cm]{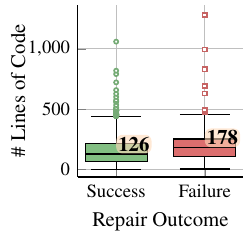}%

        \caption{Mutation generation outcome vs. input IaC complexity}
        \label{fig:subfig2}
    \end{subfigure}
    \vspace{-4mm}

    \caption{\textsc{Effectiveness analysis of multi-turn repair loops.} Box-plot of outcomes for verifiable policy and mutation generation versus input IaC complexity (\# resources, \# inter-resource relations, \# lines). In (a), median success exceeds failure; in (b), median failure exceeds success, showing policy generation improves while mutation generation worsens with complexity.}
    \Description{Effectiveness analysis of multi-turn repair loops. Two groups of box plots show outcomes for policy generation and mutation generation versus input Infrastructure as Code complexity, measured by number of resources, number of inter-resource relations, and number of lines. In the left group, corresponding to policy generation, the median success exceeds median failure. In the right group, corresponding to mutation generation, the median failure exceeds median success, indicating that policy generation improves while mutation generation worsens as complexity increases.}

    \vspace{2mm}
    \label{fig:failureanalysis}
\end{figure*}

\textbf{Experimental Setup for TerraFormer.} We build two variants of TerraFormer by fine-tuning Qwen2.5-Coder (3B and 14B) models (marked with *~and \textdagger ~in Tables~\ref{tab:iac-gen}, \ref{tab:iac-mutn}). The models are task-specific, trained for IaC generation and IaC mutation using the \textsc{TF-Gen} (Train) and \textsc{TF-Mutn} (Train) datasets, respectively. The process begins with a warm-start SFT phase to create TerraFormer (SFT), followed by RL-based fine-tuning to produce TerraFormer (SFT+RL). We implement SFT and the GRPO-based RL algorithm using Huggingface's \texttt{trl} library~\citep{vonwerra2022trl}, and train the models on 8$\times$80GB NVIDIA A100 GPUs. For the 3B model, we perform full-parameter fine-tuning with bfloat16 precision. The 14B model is fine-tuned with 4-bit NF4 quantization using LoRA~\citep{hu2022lora} applied to all linear layers in the self-attention and FFN modules, with both rank and alpha set to 16. SFT uses a learning rate of 5e-6 with cosine scheduling and a 0.05 warmup ratio, while RL uses a learning rate of 1e-6, a 0.05 warmup ratio, and DeepSpeed ZeRO-2 optimization.

\textbf{Evaluation Metrics.} We evaluate LLM-generated IaC using five metrics, covering basic syntax, functional correctness, security, and adherence to best practices. Together, these provide a holistic and practical framework, addressing the shortcomings of prior IaC evaluation approaches such as LLM Judge, CodeBLEU, and CodeBERTScore~\citep{kon2024iac} (which often overstate the quality of uncompilable code), exact string match~\citep{ganesh2023survey} (too rigid to capture multiple valid configurations), and sole reliance on static analyzers~\citep{palavalli2024using,buhler2025terrads} (which fail to capture functional correctness). The first three of our metrics, \textit{Compilability}, \textit{Deployability}, and \textit{Correctness}, form a hierarchy of increasingly strict requirements. (i) \textit{Compilability} ensures syntactic validity, confirming the configuration parses without error. (ii) \textit{Deployability} builds on this with semantic validation of cloud-specific constraints, checking whether the configuration is feasible for cloud deployment. (iii) \textit{Correctness}, the strictest of the three, demands both syntactic and semantic validity, additionally ensuring the configuration is functionally correct modulo a policy encoding the prompt’s infrastructure intent. These are evaluated using verifiers \textbf{\scriptsize FV-i}, \textbf{\scriptsize FV-ii}, and \textbf{\scriptsize FV-iii} (Section~\ref{sec:prelim}), with a Boolean pass/fail outcome. (iv) \textit{Linter Pass Rate} is a Boolean metric assessing whether a configuration complies with TFLint~\citep{tflint2024} best-practice checks. (v) \textit{Security Compliance} uses Checkov~\citep{checkov} to scan for vulnerabilities, measured as the percentage of passed checks per instance, averaged across the dataset. All metrics are evaluated using the strict \textit{pass@1} protocol, where LLMs are allowed only one generation ($k$=1) per instance. 

\textbf{Result Analysis and Ablation Studies.} Tables~\ref{tab:iac-gen} and~\ref{tab:iac-mutn} demonstrate that, for both IaC generation and mutation, accuracy tends to increase with larger LLM parameter sizes. Several broad trends are identified. \textbf{First}, among the SoTA LLMs, \textit{Sonnet 3.7 and DeepSeek-R1 achieve the highest Correctness across all three datasets}, followed by GPT-4.1. Specifically, Sonnet 3.7 attains Correctness of 35.37\% on IaC-Eval, 16.90\% on \textsc{TF-Gen} (Test), and 52.03\% on \textsc{TF-Mutn} (Test). DeepSeek-R1 shows comparable performance, with Correctness of 33.84\%, 11.65\%, and 42.03\% on the same datasets, respectively. \textbf{Second}, \textit{the distilled DeepSeek-R1 variants underperform their backbone models}. While these variants show good reasoning, they struggle with instruction-following and often hallucinate resource names or attributes. For example, on \textsc{TF-Mutn} (Test), Llama-3.3-70B-Instruct achieves Correctness 38.38\% and Deployability 77.70\%, whereas its DeepSeek-R1 distilled counterpart attains 31.49\% and 62.97\%, respectively. A similar trend appears for Qwen3-8B (Correctness 4.97\% on \textsc{TF-Gen} (Test)) versus DeepSeek-R1-0528-Qwen3-8B (2.27\%), and Llama-3.1-8B (5.26\%) versus DeepSeek-R1-Distill-Llama-8B (2.27\%). \textbf{Third}, \textit{LLMs trained on programming languages outperform similarly sized models that are not}. For example, Qwen2.5-Coder-14B achieves Correctness of 6.39\% and 35.54\% on \textsc{TF-Gen} (Test) and \textsc{TF-Mutn} (Test), respectively, compared with Qwen3-14B at 5.97\% and 30.95\%. This improvement likely stems from prior exposure to Terraform-like syntax during training. \textbf{Fourth}, \textit{LLMs perform better on IaC mutation than generation}, as mutation provides the initial configuration and clear edit instructions, reducing ambiguity. As noted in Section~\ref{sec:intro}, code mutation also better reflects real-world development practices than generating code from scratch.

For the IaC generation task on IaC-Eval, our \textbf{TerraFormer\textsuperscript{\textdagger} (SFT)} improves Qwen2.5-Coder-14B by +8.73\% in Correctness, +14.41\% in Deployability, and +16.81\% in Compilability. On \textsc{TF-Gen} (Test), the gains are +5.54\%, +26.42\%, and +20.88\%, respectively. For the IaC mutation task on \textsc{TF-Mutn} (Test), TerraFormer\textsuperscript{\textdagger} (SFT) improves Qwen2.5-Coder-14B by +8.37\%, +2.30\%, and +5.00\%. Similar trends hold for TerraFormer* (SFT) over Qwen2.5-Coder-3B. 
Our \textbf{TerraFormer\textsuperscript{\textdagger} (SFT+RL)} achieves further gains. For IaC generation on IaC-Eval, it improves Qwen2.5-Coder-14B by +15.94\%, +30.35\%, and +31.22\% in Correctness, Deployability, and Compilability, respectively; on \textsc{TF-Gen} (Test), the improvements are +11.65\%, +39.63\%, and +39.77\%. For IaC mutation on \textsc{TF-Mutn} (Test), the gains are +19.60\%, +12.71\%, and +10.54\%. Overall, TerraFormer\textsuperscript{\textdagger} (SFT+RL) achieves top rankings: on IaC-Eval (IaC generation), it ranks highest in Deployability and third in Correctness and Compilability; on \textsc{TF-Gen} (Test) (IaC generation), highest in Correctness and second in Deployability and Compilability; on \textsc{TF-Mutn} (Test) (IaC mutation), highest in Correctness and second in Deployability. 

The results show that, despite being $\approx$50$\times$ smaller, TerraFormer\textsuperscript{\textdagger} (SFT+RL) outperforms foundation LLMs such as Sonnet 3.7, DeepSeek-R1, and GPT-4.1 in Correctness on \textsc{TF-Gen} (Test) and \textsc{TF-Mutn} (Test), while ranking third on IaC-Eval. It achieves a perfect Linter Pass Rate of 100\% and the highest Security Compliance across all three benchmarks, demonstrating strong adherence to best practices. These are attributed to SFT on our rigorously verified dataset, combined with RL via a fine-grained verifier-guided reward.

\section{Quality Assessment of \textsc{TF-Gen} \& \textsc{TF-Mutn}}
\label{sec:dataset_quality}

We ensure and evaluate the quality of our datasets through multiple measures. \textbf{First}, the initial pool of Terraform configurations is sourced from real GitHub repositories, ensuring human-authored style. \textbf{Second}, for each instance $\{\mathit{{prompt}_{\text{NL}}}\allowbreak, t\allowbreak, \mathit{{policy}_{\text{FL}}}\}$ in \textsc{TF-Gen}: $t$ is deployable and $t\vdash \mathit{{policy}_{\text{FL}}}$; for each $\{\mathit{{prompt}^{\text{m}}_{\text{NL}}}, t^{\text{init}}, t^{\text{m}}, \mathit{{policy}^{\text{init}}_{\text{FL}}},\allowbreak \mathit{{policy}^{\text{m}}_{\text{FL}}}\}$ in \textsc{TF-Mutn}: both $t^{\text{init}}$ and $t^{\text{m}}$ are deployable and $t^{\text{init}}\vdash \mathit{{policy}^{\text{init}}_{\text{FL}}}$,\allowbreak $t^{\text{m}}\vdash \mathit{{policy}^{\text{m}}_{\text{FL}}}$. In this section, we analyze the total cost and effectiveness of the multi-turn repair loops. \textbf{Third}, we conduct a survey to verify that LLM-judged prompt–IaC–policy alignment (Section~\ref{sec:dataset_construction}) strongly correlates with cloud experts. \textbf{Fourth}, combining expert ratings with edit distance analysis shows that \textsc{TF-Mutn} predominantly contains medium- and high-complexity mutations.

\vspace{1mm}
\noindent\textbf{Cost and Effectiveness Analysis of Multi-Turn Repair Loops.} Given the LLM pricing of \$0.003 per 1,000 input tokens and \$0.015 per 1,000 output tokens on Amazon Bedrock, creating the dataset through repair loops costed us approximately \$15,000. Instead of using Mechanical Turk, which would have required about 30 minutes per instance at \$50/hour, totaling nearly \$5 million to curate 200,000 instances, this method is far more economical. In addition, sustaining such a large-scale human effort with cloud experts over an extended period would have been practically infeasible. 

We conduct an effectiveness analysis of the repair loops to study how policy and mutation generation (Section~\ref{sec:dataset_construction}) are affected by input configuration complexity, measured by the number of resources, inter-resource relations, and lines of code. We assess how these factors relate to success rates, defined as cases where the LLM can successfully produce a \textit{verifiable} policy or mutation within the threshold of five repair iterations. Figure~\ref{fig:failureanalysis} shows that policy generation success rises with complexity, suggesting smaller configurations lack sufficient intent or cloud primitives. In contrast, mutation success falls with complexity, as more resources and dependencies make it harder to generate deployable configurations.

\definecolor{lowimpact}{RGB}{220, 100, 100}      
\definecolor{mediumimpact}{RGB}{255, 165, 120}   
\definecolor{highimpact}{RGB}{130, 190, 130}     

\begin{figure}[t]
    \hspace{-6mm}
    \begin{subfigure}[t]{0.45\columnwidth}
        \includegraphics[width=\linewidth]{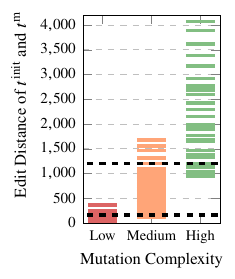}
        \vspace{-4mm}
        \caption{}
        \label{fig:survey_mutationComplexity}
    \end{subfigure}%
    \hspace{-2mm}
    \begin{subfigure}[t]{0.61\columnwidth}
        \includegraphics[width=\linewidth]{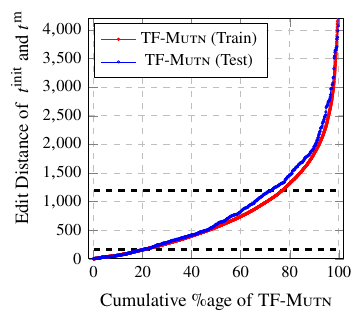}
        \vspace{-4mm}
        \caption{}
        \label{fig:wholeDataset_mutationComplexity}
    \end{subfigure}
    \vspace{-3mm}
    \caption{\textsc{Mutation Complexity in \textsc{TF-Mutn}.} (a) Correlation between edit distance and human-rated complexity for 100 survey samples; estimated thresholds for low $\mathbf{[0,160)}$, medium $\mathbf{[160,1200)}$, and high $\mathbf{[1200,\infty)}$ are shown with thick dashed lines. (b) Cumulative distribution of edit distances for all \textsc{TF-Mutn} instances by complexity, clipped at $\mathbf{4200}$ for visualization (actual maxima: Train $\mathbf{13205}$, Test $\mathbf{4619}$).}
    \Description{Mutation complexity in TF-Mutn. (a) Correlation between edit distance and human-rated mutation complexity for 100 survey samples, with estimated thresholds indicating low, medium, and high complexity ranges. (b) Cumulative distribution of edit distances for all TF-Mutn instances grouped by complexity, with values clipped at 4200 for visualization, and higher actual maxima in the training and test sets.}

    \label{fig:mutnComplexity}
\end{figure}


\vspace{1mm}
\noindent\textbf{Expert Survey on Semantic Alignment of Prompt, IaC, and Policy.} For an infrastructure instance, the policy should act as a verifier of the prompt's intent, while the target IaC should faithfully implement it. To evaluate this semantic alignment, we conduct a survey in which we uniformly sample 100 instances each from \textsc{TF-Gen} and \textsc{TF-Mutn}. We then partition the instances into 20 groups of 10 and assign each group to a rater with expertise in cloud infrastructures. For each \textsc{TF-Gen} instance, we pose two questions: 
(a) \textbf{Policy$\boldsymbol{\leftrightarrow}$Prompt}: ``\textit{To what extent does $\mathit{{policy}_{\text{FL}}}$ capture the intent of $\mathit{{prompt}_{\text{NL}}}$ (treating the former as a test-case for the latter)?}'' and 
(b) \textbf{Prompt$\boldsymbol{\leftrightarrow}$IaC}: ``\textit{To what extent is the intent of $\mathit{{prompt}_{\text{NL}}}$ correctly realized by $t$?}''. For \textsc{TF-Mutn}, we define $\Delta$IaC = diff$\bigl(t^{\text{init}}, t^{\text{m}}\bigr)$ and $\Delta$Policy = diff$\bigl(\mathit{policy^{\text{init}}_{\text{FL}}}, \mathit{policy^{\text{m}}_{\text{FL}}}\bigr)$, and ask:
(a) \textbf{$\boldsymbol{\Delta}$Policy$\boldsymbol{\leftrightarrow}$$\boldsymbol{\Delta}$IaC}: ``\textit{To what extent does $\Delta$Policy reflect the $\Delta$IaC changes?}'' and 
(b) \textbf{$\boldsymbol{\Delta}$IaC$\boldsymbol{\leftrightarrow}$Prompt}: ``\textit{To what extent does $\Delta$IaC match $\mathit{{prompt}^{\text{m}}_{\text{NL}}}$'s intent?}''.
Responses use a three-point scale: \textit{Significant}, \textit{Moderate}, and \textit{Slight}. To address subjectivity, three raters independently evaluated 25 instances each from \textsc{TF-Gen} and \textsc{TF-Mutn}, and we measure inter-rater agreement using Gwet’s AC1~\citep{gwet2001handbook}, which provides stable estimates even for skewed response distributions.

The results, summarized in Figure~\ref{fig:humanAnnot}, show strong semantic alignment. For the two \textsc{TF-Gen} questions, 91\% and 96\% of raters rate the alignment as \textit{Significant}, and for \textsc{TF-Mutn}, 96\% and 99\%. IRA remains high, with Gwet's AC1 coefficients of 89.87\%, 91.71\%, 94.47\%, and 97.30\% across the four questions. Furthermore, all instances in the datasets and effectively all in the survey are approved by the LLM judge during dataset construction. This ensures that the results directly reflect cloud experts' agreement with the LLM judge, demonstrating its effectiveness in validating semantic alignment.



\vspace{1mm}
\noindent\textbf{Mutation Complexity in \textsc{TF-Mutn}.} From the expert survey, we measure the mutation complexity by asking raters to first identify differences between ($t^{\text{init}}$ and $t^{\text{m}}$), then rate $\boldsymbol{\Delta}$\textbf{IaC complexity}: ``\textit{How challenging would it be for a cloud expert to mutate $t^{\text{init}}$ into $t^{\text{m}}$ given the significance of differences?}''. Responses use a three-point scale: Low, Medium, High. Although subjective, the ratings show clear patterns: minor edits (e.g., tag or feature-toggle updates) were rated \textit{low}; additions or adjustments of resources without changing the overall architecture were rated \textit{medium}; and substantial redesigns (e.g., adding multiple components or restructuring interactions) were rated \textit{high}. To quantify this, we compute the Levenshtein (edit) distance between $t^{\text{init}}$ and $t^{\text{m}}$ for all survey samples and correlate them with expert ratings (Figure~\ref{fig:survey_mutationComplexity}). Thresholds minimizing disagreement suggest distances below 160 generally correspond to low complexity, 160–1200 to medium, and above 1200 to high.

Extending this analysis, we compute the edit distances for all 51,776 instances in \textsc{TF-Mutn} (Train) and 740 in \textsc{TF-Mutn} (Test), and plot the cumulative distribution (Figure~\ref{fig:wholeDataset_mutationComplexity}). Using survey-derived thresholds, mutations are categorized by complexity: in \textsc{TF-Mutn} (Train), 20.35\% are low-complexity mutations, 56.82\% medium, and 22.83\% high; in \textsc{TF-Mutn} (Test), 20\% are low, 52.57\% medium, and 27.43\% high. Both datasets thus contain a balanced mix, with medium and high predominating, providing a diverse and challenging set for validating and training IaC mutation tools.

\section{Conclusion}
\label{sec:concl}
We present \emph{TerraFormer}, a framework for automated IaC generation and mutation using LLMs fine-tuned with policy-guided verifier feedback. It addresses key IaC automation challenges: scarce high-quality datasets, difficulty generating syntactically and functionally correct IaC, and lack of reliable evaluation metrics. It is also among the first to tackle IaC mutation. Our automated data curation pipeline produces large NL-to-IaC datasets, \textsc{TF-Gen} and \textsc{TF-Mutn}, via multi-stage verification and iterative LLM self-correction. Combining SFT with RL guided by fine-grained verifier feedback, TerraFormer significantly improves syntax correctness, deployability, policy compliance, and security over its base LLM. Evaluations against 17 state-of-the-art LLMs, including models nearly 50$\times$ larger such as Sonnet 3.7, DeepSeek-R1, and GPT-4.1, show TerraFormer outperforms them on \textsc{TF-Gen} (Test) and \textsc{TF-Mutn} (Test), achieves third-best on IaC-Eval, and attains highest best-practices and security compliance rates. These results demonstrate that verifier-guided fine-tuning on formally verified datasets produces more intent-aligned, deployable IaC. Overall, TerraFormer provides a scalable foundation for automating cloud infrastructure and other code generation tasks, highlighting the potential of combining LLM reasoning with formal verification for IaC automation.



\clearpage
\appendix

\section{Reproducibility}
\label{sec-appendix:reprod}

All experiments in this work, including dataset construction, model training (both supervised fine-tuning and reinforcement learning), inference, and evaluation, are performed using Python 3.12.10 and Terraform v1.12.0. Our full codebase, including scripts for dataset generation, model fine-tuning, and inference across both GPU- and API-based setups, can be shared upon request. 


\section{Description of LLM Prompts}
\label{sec-appendix:prompt_description}

\lstset{
    basicstyle=\ttfamily\tiny,
    breaklines=true,
    breakindent=0pt,        
    frame=single,
    backgroundcolor=\color{gray!10}
}

\noindent \textbf{Prompt Template for Multi-Turn Repair Loop.} 
We present the prompt used in step-3 of Figure~\ref{fig:archDiagram-seed}, where 
\textbf{\scriptsize FV-ii} (\texttt{terraform plan}) acts as the verifier. 
Given a Terraform configuration, \textbf{\scriptsize FV-ii} produces an error certificate if the configuration fails the deployability test. This certificate guides iterative repairs in a multi-turn loop, with each round updating both the certificate and the configuration until the configuration passes the deployability test. Similar prompt templates are employed for repair loops 
with other verifiers, \textbf{\scriptsize FV-i} and \textbf{\scriptsize FV-iii}, 
with adjustments in task description and requirements.

\vspace{1mm}
\begin{lstlisting}
You are an expert Infrastructure-as-Code (IaC) developer with deep expertise in Terraform.

You will be given:
(a) A Terraform configuration written in HCL, intended for Terraform v1.12.0. This code currently fails the terraform plan command.
(b) The error message output from running terraform plan.

Your task is to:
(a) Clearly explain the cause of the error in your own words.
(b) Describe how you will fix the issue.
(c) Provide a fully corrected version of the Terraform configuration (in HCL) that ensures terraform plan completes successfully.

Requirements:
(i) If the error indicates "timed out", it is possibly because Terraform waits for interactive input to fill a variable with no default value.
    Ensure that you add reasonable default values for all input variables for non-interactive execution.
(ii) You are permitted to modify any aspect of the Terraform code, including adding or removing entire blocks, arguments, or resources. Apply changes as rigorously as necessary to ensure that terraform plan completes successfully.
(iii) You must ensure compatibility with Terraform version 1.12.0.
(iv) You may remove problematic elements such as assume_role blocks or profile = "admin-1" if they can cause permission errors.
(v) Terraform will rely on default credentials (e.g., EC2 instance metadata or environment variables), so explicit credential configuration should be removed if problematic.
(vi) You are only allowed to change the Terraform code. Do not assume access to other configuration files or systems.

Here are three examples of correct Terraform HCL codes:

### Example-1
```hcl
{TF_example1}
```

### Example-2
```hcl
{TF_example2}
```

### Example-3
```hcl
{TF_example3}
```

Here is the incorrect configuration (in Terraform HCL):
<incorrect_terraform_config>
```hcl
{config}
```
</incorrect_terraform_config>

Here is the error message obtained by running terraform plan command or a timeout notice if the command does not complete within 30 seconds:
<error_message>
{error_message_TFplan}
</error_message>

Return the correct Terraform configuration within the following tags. Do not return empty code.
<corrected_terraform_config>
(Your entire Terraform configuration goes here)
</corrected_terraform_config>
\end{lstlisting}

\newpage
\noindent \textbf{Prompt Template for few-shot inference of IaC generation.} 
The following prompt is used to obtain the results of all the LLM-based tools reported in Table~\ref{tab:iac-gen} for the IaC generation task.

\begin{lstlisting}
You are an expert Infrastructure-as-Code (IaC) developer with deep expertise in Terraform.

Your task is to:
Given an user prompt, generate a **single**, **fully self-contained**, and **valid** Terraform configuration written in HCL. Your configuration must:
- Satisfy the intent of the user prompt.
- Be compatible with **Terraform v1.12.0**.
- Pass both `terraform validate` and `terraform plan`.
- Include a valid `provider` block.
- Contain no undeclared variables or references.
- Avoid the use of `assume_role`, custom `profile` values, or external dependencies.
- Use only the following providers: `aws`, `random`, `null`, `local`, `template`, `tls`, `time`, `external`, `http`, `archive`, `docker`, `terraform`.

Here are a few examples:

### Example-1 
Prompt: {prompt_example1}
Configuration: 
```hcl
{TF_gen_example1}
```
...

Here is the **actual** user prompt:
<user_prompt>
{request}
</user_prompt>

Now respond to the actual user prompt by returning **one single** Terraform configuration. Do not repeat or revise the configuration. Importantly, enclose the final configuration inside the following tags:

<final_terraform_config>
(Provide your entire Terraform configuration within these tags)
</final_terraform_config>
\end{lstlisting}

\vspace{2mm}
\noindent \textbf{Prompt Template for few-shot inference of IaC mutation.} The following prompt is used to obtain the results of all the LLM-based tools reported in Table~\ref{tab:iac-mutn} 
for the IaC mutation task.
\begin{lstlisting}
You are an expert Infrastructure-as-Code (IaC) developer with deep expertise in Terraform.

Your task is to:
Given an existing Terraform configuration and an user prompt requesting changes, generate a **single**, **modified** Terraform configuration written in HCL that is **fully self-contained** and **valid**. Your configuration must:
- Be a modified version of the existing Terraform configuration.
- Satisfy the intent of the changes requested in the user prompt.
- Be compatible with **Terraform v1.12.0**.
- Pass both `terraform validate` and `terraform plan`.
- Include a valid `provider` block.
- Contain no undeclared variables or references.
- Avoid the use of `assume_role`, custom `profile` values, or external dependencies.
- Use only the following providers: `aws`, `random`, `null`, `local`, `template`, `tls`, `time`, `external`, `http`, `archive`, `docker`, `terraform`.

Here are a few examples:

### Example-1 
Initial Configuration:
```hcl
{TF_init_example1}
```
Prompt: {prompt_example1}
Mutated Configuration:
```hcl
{TF_mutn_example1}
```
...

Here is the **initial Terraform configuration**:
<initial_terraform_config>
```hcl
{TF_init}
```
</initial_terraform_config>

Here is the **user prompt** requesting changes:
<user_prompt>
{prompt}
</user_prompt>

Now respond by returning **a single**, **modified** Terraform configuration. Importantly, enclose the final configuration inside the following tags:

<mutated_terraform_config>
(Provide your entire modified Terraform configuration within these tags)
</mutated_terraform_config>
\end{lstlisting}

\clearpage
\bibliographystyle{ACM-Reference-Format}
\bibliography{references}

\end{document}